%% file: higherweight5pt.tex
\RequirePackage{pdf14}

\documentclass[a4paper,11pt]{article}
\pdfoutput=1 

\usepackage[outline]{contour}
\newcommand*{\fancy}[1]{{\color{black}\contour{gray}{#1}}}
\usepackage{array,setspace,mathrsfs,amsfonts,amsmath,yfonts,dsfont,bbm,colonequals,amscd}
\usepackage{relsize,suffix,mathtools,cancel,bbm,tikz-cd}
\usepackage{subcaption}
\usepackage{bbold}

\usepackage{jheppub} 

\usepackage[T1]{fontenc} 
\usepackage{subcaption}
\graphicspath{ {graphs/} }

\usepackage{xcolor}

\abstract{We continue to explore the bootstrap approach to five-point correlation functions for IIB supergravity on $AdS_5\times S^5$. Building on the result of \cite{Goncalves:2019znr}, we develop an improved algorithm that allows us to more efficiently compute correlators of higher Kaluza-Klein modes. The new method uses only factorization and a superconformal twist, and is entirely within Mellin space where the analytic structure of holographic correlators is simpler. Using this method, we obtain in a closed form all five-point functions of the form $\langle pp222\rangle$, extending the earlier result for $p=2$. As a byproduct of our analysis, we also obtain explicit results for spinning four-point functions of higher Kaluza-Klein modes.
}

\input{sections/titlepage}

\begin{document}
\maketitle
\flushbottom

\input{sections/introduction}
\input{sections/scfkine}

\input{sections/Mellinrep}
\input{sections/bootstrapMellin}

\input{sections/discussion}

\acknowledgments
V.G. was supported by Simons Foundation grants \#488637 (Simons collaboration on the non-perturbative bootstrap), FAPESP grant 2015/14796- 7 and by the Coordenacao de Aperfeicoamento de Pessoal
de Nivel Superior - Brasil (CAPES) - Finance Code 001. Centro de Fisica do Porto is partially funded by Fundacao para a Ciencia e Tecnologia (FCT) under the grant UID04650-FCUP. The work of C.M. has been supported in part by Istituto Nazionale di Fisica Nucleare (INFN) through the ``Gauge and String Theory'' (GAST) research project. X.Z. is supported by funds from University of Chinese Academy of Sciences (UCAS), funds from the Kavli Institute for Theoretical Sciences (KITS), the Fundamental Research Funds for the Central Universities, and the NSFC Grant No. 12275273. The work of X.Z. carried out at Princeton was supported in part by the Simons Foundation Grant No. 488653. JVB is funded by FCT with fellowship DFA/BD/5354/2020, co-funded by Norte Portugal Regional Operational Programme
(NORTE 2020), under the PORTUGAL 2020 Partnership Agreement, through the European
Social Fund (ESF).

\appendix
\input{sections/appendixKinematics}
\input{sections/appendixMellin}

\bibliography{refs} 
\bibliographystyle{utphys}

\end{document}

%% file: sections/titlepage.tex
\title{\boldmath \LARGE
Kaluza-Klein Five-Point Functions from AdS$_5$$\times$S$^5$ Supergravity 
 }

\author[a,d]{Vasco Gon\c{c}alves,}
\author[b]{Carlo Meneghelli,}
\author[c]{Raul Pereira,}
\author[d]{Joao Vilas Boas,}
\author[e,f]{Xinan Zhou}

\affiliation[a]{ICTP South American Institute for Fundamental Research, IFT-UNESP, \\S\~ao Paulo, SP Brazil 01440-070}
\affiliation[b]{ Dipartimento SMFI, Universit\`a di Parma,
Viale G.P.~Usberti 7/A, 43121, Parma, PR, Italy}
\affiliation[c]{School of Mathematics and Hamilton Mathematics Institute, Trinity College Dublin,\\
Dublin 2, Ireland}
\affiliation[d]{Centro de Fisica do Porto e Departamento de Fisica e Astronomia, Faculdade de Ciencias da Universidade
do Porto, Rua do Campo Alegre 687, 4169-007, Porto, Portugal}
\affiliation[e]{Kavli Institute for Theoretical Sciences,
University of Chinese Academy of Sciences, \\Beijing 100190, China}
\affiliation[f]{Princeton Center for Theoretical Science, Princeton University, Princeton, NJ 08544, USA}

\emailAdd{vasco.dfg@gmail.com}
\emailAdd{carlo.meneghelli@unipr.it}
\emailAdd{raul@maths.tcd.ie}
\emailAdd{joaomiguelvb@gmail.com}
\emailAdd{xinan.zhou@ucas.ac.edu}


%% file: sections/introduction.tex
\section{Introduction}\label{Sec:introduction}
Recent years have seen significant progress in computing holographic correlators, which are key objects for exploring and exploiting the AdS/CFT correspondence. Traditionally, holographic correlators are computed by diagrammatic expansions in AdS. Such a method works in principle. However, in practice, it requires the precise knowledge of the exceedingly complicated effective Lagrangians and is extremely cumbersome to use. Therefore, for almost twenty years the diagrammatic approach led to only a handful of explicit results. The new developments, on the other hand, are based a totally different strategy which relies on new principles. This is the bootstrap approach initiated in \cite{Rastelli:2016nze,Rastelli:2017udc}, which eschews the explicit details of the effective Lagrangian altogether. The new approach works directly with the holographic correlators and uses superconformal symmetry and consistency conditions to fix these objects. The bootstrap strategy has produced an array of impressive results.\footnote{See \cite{Bissi:2022mrs} for a review.} For example, at tree level general four-point functions for $\frac{1}{2}$-BPS operators with arbitrary Kaluza-Klein (KK) levels have been computed in all maximally superconformal theories \cite{Rastelli:2016nze,Rastelli:2017udc,Alday:2020lbp,Alday:2020dtb}, as well as in theories with half the amount of maximal superconformal symmetry \cite{Rastelli:2019gtj,Giusto:2020neo,Alday:2021odx}. Note that these  general results are all in the realm of four-point functions. Higher-point functions still mostly remain {\it terra incognita}. In fact, only two five-point functions have been computed in the literature for IIB supergravity on $AdS_5\times S^5$ \cite{Goncalves:2019znr} and SYM on $AdS_5\times S^3$ \cite{Alday:2022lkk} respectively, and both for the lowest KK modes only. 

 However, studying higher-point holographic correlator is of great importance. Firstly, higher-point correlators allow us to extract new CFT data which is not included in four-point functions. For example, the OPE coefficient of two double-trace operators and one single-trace operator can only be obtained from a five-point function. Moreover, via the AdS unitarity method \cite{Aharony:2016dwx} higher-point correlators are also necessary ingredients for constructing higher-loop correlators. Secondly, via the AdS/CFT correspondence holographic correlators correspond to on-shell scattering amplitudes in AdS. Recently, there has been a lot of progress in finding AdS generalizations of flat-space properties \cite{Farrow:2018yni,Armstrong:2020woi,Albayrak:2020fyp,Alday:2021odx,Jain:2021qcl,Zhou:2021gnu,Diwakar:2021juk,Alday:2022lkk,Cheung:2022pdk,Herderschee:2022ntr,Drummond:2022dxd,Bissi:2022wuh,Armstrong:2022jsa,Lee:2022fgr,Li:2022tby}. As we know from flat space, many remarkable properties of amplitudes are only visible at higher multiplicities. To further explore the analogy between holographic correlators and scattering amplitudes it is necessary to go to higher points. Finally, it has been observed in \cite{Caron-Huot:2018kta} that a ten dimensional hidden conformal symmetry is responsible for organizing all tree-level four-point functions for IIB supergravity on $AdS_5\times S^5$. The nature of this hidden structure is still elusive. It is an interesting question whether the 10d hidden symmetry is just a curiosity for four points or it persists even at higher points.

For these reasons, in this paper we continue to explore the bootstrap strategy for computing higher-point correlators. In particular, we will focus on computing the five-point functions of the form $\langle pp222\rangle$ for IIB supergravity in $AdS_5\times S^5$, where three of the operators have the lowest KK level but the other two have arbitrary KK level $p$. Our strategy will be similar to that of \cite{Goncalves:2019znr}, which computed the $p=2$ case, but with important differences. In \cite{Goncalves:2019znr}, the starting point is an ansatz in position space which is a linear combination of all possible Witten diagrams with unfixed coefficients. To fix the coefficients, one imposes various constraints from superconformal symmetry and consistency conditions. These includes factorization in Mellin space \cite{Goncalves:2014rfa}, the chiral algebra constraint \cite{Beem:2013sza} and the Drukker-Plefka twist \cite{Drukker:2009sf}. The first constraint is the consistency condition for decomposing the five-point function into four-point functions and three-point functions at its singularities. The second and the third conditions come from superconformal symmetry and are the statement that the appropriately twisted five-point function becomes topological. Although these conditions uniquely fix the $p=2$ five-point function, the strategy of \cite{Goncalves:2014rfa} suffers from a few drawbacks which make it difficult to apply efficiently to correlators with higher KK levels. Firstly, computing the higher-point Witten diagrams in the ansatz is a nontrivial task.  In particular, simplifications used in \cite{Goncalves:2014rfa} for computing $p=2$ diagrams no longer exist for $p>2$ and the analysis is in general more complicated. Secondly, the three constraints were implemented in difference spaces, which makes the algorithm less efficient. Factorization is most convenient in Mellin space.  However, the chiral algebra constraint and the Drukker-Plefka twist were implemented in the original position space. The position space implementation requires computing explicitly a set of five-point contact diagrams, {\it i.e.,} $D$-functions, to which the ansatz reduces. As was shown in \cite{Goncalves:2014rfa}, these $D$-functions can further be expressed in terms of one-loop box diagrams which can be written as ${\rm Li}_2$ and logarithms. But the complexity of the expression for each $D$-function is determined by its total external conformal dimensions. For $\langle pp222\rangle$ five-point functions, the sum of dimensions grows linearly with respect to $p$. Therefore, it soon becomes computationally very expensive for large enough $p$.

We overcome these difficulties by proposing a new algorithm. It relies on the key observation that a more careful analysis of the Mellin factorization condition together with the Drukker-Plefka twist allow us to completely fix the five-point correlators without using the chiral algebra constraint. Although computing Witten diagrams is difficult in position space, formulating the ansatz in Mellin space is straightforward thanks to their simplified analytic structure in Mellin space. This is further aided by a new pole truncation phenomenon which keeps the number of poles fixed irrespective of the KK levels. As a result, we can write down the ansatz for the Mellin amplitude for general $p$. Moreover, we find a way to implement the Drukker-Plefka twist directly in Mellin space. Therefore, we can perform the bootstrap entirely within Mellin space without ever taking the position space detour. This allows us to compute the five-point $\langle pp222\rangle$ Mellin amplitudes for arbitrary $p$ in a closed form. Although in this paper we focused on this particular family of correlators, the strategy applies straightforwardly to more general five-point functions.

The rest of the paper is organized as follows. In Sec. \ref{Sec:scfkine} we discuss the superconformal kinematics of the five-point functions. In particular, we will introduce the Drukker-Plefka twist. In Sec. \ref{Sec:Mellin} we review the Mellin space formalism and the factorization of Mellin amplitudes. We also explain how to implement the Drukker-Pleka twist in Mellin space. We bootstrap the five-point functions in Sec. \ref{Sec:bootstrapMellin} and give the general formula for the $\langle pp222\rangle$ Mellin amplitudes. In Sec. \ref{Sec:bootstrapposition} we also comment on how to perform the bootstrap in position space. We conclude in Sec. \ref{Sec:discussion} with an outlook for future directions. Technical details are contained in the two appendices. In Appendix \ref{Sec:higherkkSmult} we explain how to compute spinning four-point functions which are needed for factorizing the five-point functions. In Appendix \ref{Sec:rSymmetryGluing} we discuss how to glue together the R-symmetry dependence when performing factorization. For reader's convenience, we also included a Mathematica notebook with the arXiv submission which contains various explicit results. 

%% file: sections/scfkine.tex
\section{Superconformal kinematics of five-point functions}\label{Sec:scfkine}

We consider the correlation functions of the super primaries of the $\frac{1}{2}$-BPS multiplets. These are scalar operators $\mathcal{O}_k^{I_1\ldots I_k}$ with $I=1,\ldots,6$, $k=2,3,\ldots$, transforming in the rank $k$ symmetric traceless representation of the $SO(6)$ R-symmetry group. Their conformal dimensions are protected by supersymmetry and are determined by the R-symmetry representation $\Delta=k$. Via the AdS/CFT correspondence, they are dual to scalar fields in AdS with KK level $k$ and are usually referred to as the super gravitons. A convenient way to keep track of the R-symmetry information is to contract the indices with null polarization vectors
\begin{equation}
\mathcal{O}_k(x;t)=\mathcal{O}_k^{I_1\ldots I_k}t_{I_1}\ldots t_{I_k}\;,\quad t\cdot t=0\;.
\end{equation}
Our main target in this paper is the following five-point correlator
\begin{equation}
G_p(x_i;t_i)=\langle \mathcal{O}_p(x_1;t_1)\mathcal{O}_p(x_2;t_2) \mathcal{O}_2(x_3;t_3) \mathcal{O}_2(x_4;t_4) \mathcal{O}_2(x_5;t_5) \rangle\;.\label{eq:fivePtkk222def}
\end{equation}
More precisely, we will compute the leading connected contribution which is of order $1/N^3$ and corresponds to tree-level scattering in AdS. The disconnected piece factorize into a three-point function and a two-point function, and is protected because the lower-point functions are.

Symmetry imposes strong constraints on the form the correlator. For example, conformal symmetry allows us to write the five-point function as a function of five conformal cross ratios after extracting an overall kinematic factor\footnote{We are using a different, but equivalent, set of cross ratios here compared to \cite{Goncalves:2019znr}. These new cross ratios have appeared before in \cite{Bercini:2020msp}. One reason why these variables are nice is that it is possible to associate some $x_{ij}^2$ to  $u_k$, for example $x_{12}$ only appears in $u_1$. Another interesting property is that they are cyclically related to each other.    }
\begin{equation}
u_1 = \frac{x_{12}^2 x_{35}^2}{x_{13}^2 x_{25}^2}, \ \ u_2=\frac{x_{14}^2 x_{23}^2}{x_{13}^2 x_{24}^2}, \ \ u_3=\frac{x_{25}^2 x_{34}^2}{x_{24}^2 x_{35}^2}, \ \ u_4=\frac{x_{13}^2 x_{45}^2}{x_{14}^2 x_{35}^2}, \ \ u_5=\frac{x_{15}^2 x_{24}^2}{x_{14}^2 x_{25}^2}\label{eq:crossratiosPos}
\end{equation}
where we have defined $x_{ij}=x_i-x_j$. Similarly, extracting a kinematic factor also allows us to express the R-symmetry dependence as a function of the following five R-symmetry cross ratios
\begin{equation}
\sigma_1 = \frac{t_{12}t_{35}}{t_{13}t_{25}}, \ \ \sigma_2=\frac{t_{14}t_{23}}{t_{13}t_{24}}, \ \ \sigma_3=\frac{t_{25}t_{34}}{t_{24}t_{35}}, \ \ \sigma_4=\frac{t_{13}t_{45}}{t_{14}t_{35}}, \ \ \sigma_5=\frac{t_{15}t_{24}}{t_{14}t_{25}}\label{eq:crossratiosRCharge}
\end{equation}
where we have introduced the shorthand notation $t_{ij}=t_i\cdot t_j$. However, there is more we can say about the R-symmetry dependence. Since the polarization vectors $t_i$ are just multiplied to saturate the R-symmetry indices, they must appear in $G_p$ with positive powers. Therefore, $G_p$ must be a collection of monomials of the form $\prod_{i<j}t_{ij}^{a_{ij}}$, with the conditions
\begin{equation}
a_{ij}=a_{ji}\geq 0\;,\quad \sum_{j\neq i}a_{ij}=k_i\;,
\end{equation}
where $k_1=k_2=p$, $k_3=k_4=k_5=2$ are the weights of the external operators. Note the number of these monomials is finite and we will refer to them as different R-symmetry structures. In Section \ref{Subsec:Rsymm}, we will explicitly write down these structures. 

The considerations so far have only used the bosonic symmetries in the full superconformal group. The dependence on the spacetime variables $x_{ij}^2$ and on the R-symmetry variables $t_{ij}$ are not related. However, the fermionic generators in the superconformal group will impose further constraints which correlate the $x_{ij}^2$ and $t_{ij}$ dependence. For five-point functions, a thorough analysis the full consequence of the fermionic symmetries has not been performed in the literature. However, two classes of such constraints are known. The first is the chiral algebra construction \cite{Beem:2013sza} which constrain the five-point function when all the operators are inserted on a two dimensional plane. The other is the Drukker-Plefka twist \cite{Drukker:2009sf} which imposes constraints on the correlator with generic insertion positions. In this paper, we will only need the latter. We will review these conditions in Section \ref{Subsec:DrukkerPlefkatwist}.

\subsection{R-symmetry}\label{Subsec:Rsymm}

\begin{figure}[h]
\centering
\includegraphics[width=\textwidth]{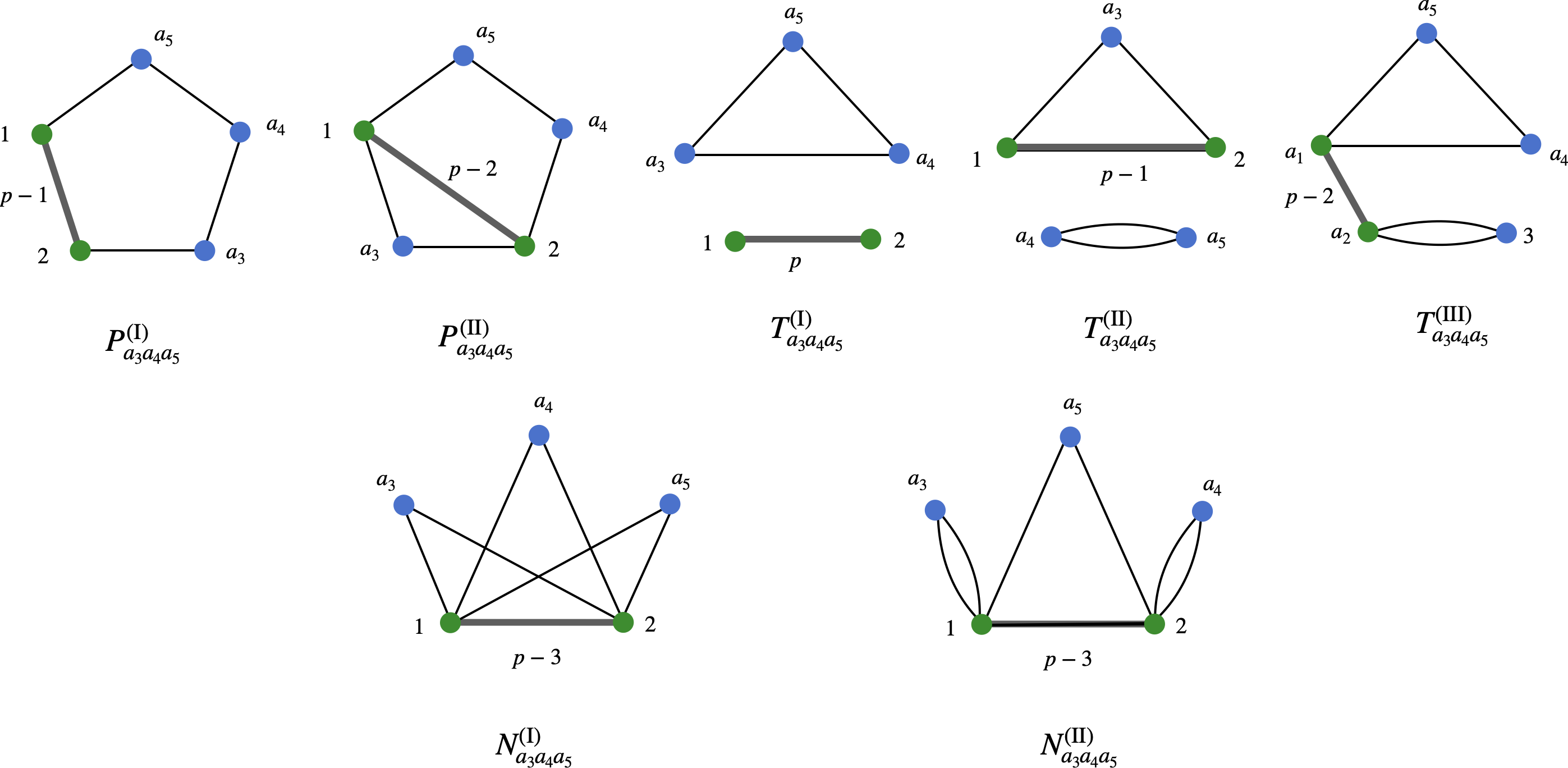}
\caption{Inequivalent R-symmetry structures in the $\langle pp222\rangle$ five-point function. Here $(a_1,a_2)$ is $(1,2)$ or $(2,1)$ and  $(a_3,a_4,a_5)$ can be any permutation of $(3,4,5)$. Each thin line represents a single contraction. The thick line represents the multi-contraction $t_{12}^a$ with the power $a$ given by the number next to the line. The R-symmetry structures in the first row have counterparts in the $\langle22222\rangle$ five-point correlator. For $\langle pp222\rangle$ they are simply obtained by multiplying the $p=2$ structures with $t_{12}^{p-2}$. The R-symmetry structures in the second row are new and do not appear in $\langle22222\rangle$.}
    \label{fig:Rsymmstructures}
\end{figure}

A systematic way to enumerate the R-symmetry structures of the $\langle pp222\rangle$ five-point function is to consider the Wick contractions. Different Wick contractions are illustrated in Fig. \ref{fig:Rsymmstructures} and the corresponding R-symmetry structures are explicitly given by
\begin{equation}
\begin{split}
P^{({\rm I})}_{a_3a_4a_5}={}&t_{12}^{p-1}t_{2a_3}t_{a_3a_4}t_{a_4a_5}t_{1a_5}\;,\\
P^{({\rm II})}_{a_3a_4a_5}={}&t_{12}^{p-2}t_{1a_3}t_{2a_3}t_{2a_4}t_{a_4a_5}t_{1a_5}\;,\\
T^{({\rm I})}_{a_3a_4a_5}={}&t_{12}^pt_{a_3a_4}t_{a_4a_5}t_{a_3a_5}\;,\\
T^{({\rm II})}_{a_3a_4a_5}={}&t_{12}^{p-1}t_{2a_3}t_{1a_3}t_{a_4a_5}^2\;,\\
T^{({\rm III})}_{a_1a_2a_3a_4a_5}={}&t_{a_1a_2}^{p-2}t_{a_1a_4}t_{a_4a_5}t_{1a_5}t_{a_2a_3}^2\;,\\
N^{({\rm I})}_{a_3a_4a_5}={}&t_{12}^{p-3}t_{1a_3}t_{1a_5}t_{2a_3}t_{2a_5}t_{1a_4}t_{2a_4}\;,\\
N^{({\rm I})}_{a_3a_4a_5}={}&t_{12}^{p-3}t_{1a_3}^2t_{2a_4}^2t_{1a_5}t_{2a_5}\;.
\end{split}
\end{equation}
Here $(a_1,a_2)$ is $(1,2)$ or $(2,1)$ and  $(a_3,a_4,a_5)$ can be any permutation of $(3,4,5)$. The Wick contractions in the first row of Fig. \ref{fig:Rsymmstructures} exist for all $p\geq 2$ while the second row are only possible when $p\geq 3$. This is a new phenomena that arises at the level of five-point functions and should be contrasted with the four-point function case. In the four-point function $\langle pp22\rangle$, the number of Wick contractions is the same irrespective of the Kaluza-Klein weight $p$.\footnote{In fact, this is true even in the more general case $\langle pqrs \rangle$ as long as the extremality $E$ of the correlator remains the same. Here extremality is defined as $E=s-p-q-r$ and we have assumed that $s$ is the largest weight of them.} 

For $p=2$, all the five points are on the same footing and there is no distinction between $P^{({\rm I})}_{a_3a_4a_5}$, $P^{({\rm II})}_{a_3a_4a_5}$ and among $T^{({\rm I})}_{a_3a_4a_5}$, $T^{({\rm II})}_{a_3a_4a_5}$, $T^{({\rm III})}_{a_1a_2a_3a_4a_5}$. Multiplying them by $t_{12}^{p-2}$ gives the corresponding structures when $p>2$. Note that even when $p\geq 3$, some of these R-symmetry structures in Fig. \ref{fig:Rsymmstructures}  still have residual symmetries and are invariant under certain permutations of $\{a_3,a_4,a_5\}$. For example, $T^{({\rm I})}_{a_3a_4a_5}=T^{({\rm I})}_{a_4a_3a_5}=T^{({\rm I})}_{a_3a_5a_4}$ and $T^{({\rm II})}_{a_3a_4a_5}=T^{({\rm II})}_{a_3a_5a_4}$. We choose the independent R-symmetry structures to be 
\begin{equation}\label{indepRstructures}
\begin{split}
{}&P^{({\rm I,II})}_{a_3a_4a_5}:\quad\quad (a_3,a_4,a_5)\in\{(3,4,5),(3,5,4),(4,3,5),(4,5,3),(5,3,4),(5,4,3)\}\;,\\
{}& T^{({\rm I})}_{a_3a_4a_5}:\quad\quad (a_3,a_4,a_5)\in\{(3,4,5)\}\;,\\
{}& T^{({\rm II})}_{a_3a_4a_5}:\quad\quad (a_3,a_4,a_5)\in\{(3,4,5),(4,3,5),(5,3,4)\}\;,\\
{}& T^{({\rm III})}_{a_1a_2a_3a_4a_5}:\;\;\; (a_1,a_2,a_3,a_4,a_5)\in\{(1,2,3,4,5),(1,2,4,3,5),(1,2,5,3,4),(2,1,3,4,5),\\
{}&\quad\quad\quad\quad\quad\quad\quad(2,1,4,3,5),(2,1,5,3,4)\}\;,\\
{}& N^{({\rm I})}_{a_3a_4a_5}:\quad\quad (a_3,a_4,a_5)\in\{(3,4,5)\}\;,\\
{}& N^{({\rm II})}_{a_3a_4a_5}:\quad\quad (a_3,a_4,a_5)\in\{(3,4,5),(3,5,4),(4,3,5),(4,5,3),(5,3,4),(5,4,3)\}\;.
\end{split}
\end{equation}
This gives in total 29 independent R-symmetry structures. When $p=2$, $N^{({\rm I})}_{a_3a_4a_5}$ and $N^{({\rm II})}_{a_3a_4a_5}$ do not exist and we have 22 structures. 

\subsection{Drukker-Plefka twist and chiral algebra}\label{Subsec:DrukkerPlefkatwist}
A highly nontrivial constraint from superconformal symmetry is given by the topological twist discovered in \cite{Drukker:2009sf}, which we will refer to as the Drukker-Plefka twist. In  \cite{Drukker:2009sf}, it was found that when the operators have the following position-dependent polarization vectors (commonly referred to as a twist)
\begin{equation}
\bar{t}_i=(i x_i^1,i x_i^2,i x_i^3,i x_i^4,\frac{i}{2}(1-(x^\mu)^2),\frac{1}{2}(1+(x^\mu)^2))\;,
\end{equation}
the twisted correlator preserves certain nilpotent supercharge. The twisted operators are in its cohomology. More importantly, the translations of operators while keeping the polarizations twisted are exact. It then follows that the twisted correlators are topological, {\it i.e.}, independent of the insertion locations
\begin{equation}\label{twistGp}
G_p(x_i;\bar{t}_i)={\rm constant}\;.
\end{equation}
Note that in terms of the variables $x_{ij}^2$ and $t_{ij}$, the twist condition can also be written as $t_{ij}=x_{ij}^2$.

Let us also mention another twist for contrast, namely the chiral algebra \cite{Beem:2013sza}. However, we will not exploit this twist in this paper. The chiral algebra twist requires that all the operators are inserted on a two dimensional plane. The coordinates therefore can be parameterized by the complex coordinates $z$, $\bar{z}$. Furthermore, the polarization vectors need to be restricted to be four dimensional 
\begin{equation}
t_i=(t_i^\mu,0,0)\;,\quad \mu=1,2,3,4\;,
\end{equation}
where $t^\mu$ can be written in terms of two-component spinors
\begin{equation}
    t_i^\mu=\sigma^\mu_{\alpha\dot{\alpha}}v^\alpha\bar{v}^{\dot{\alpha}}\;.
\end{equation}
Using the rescaling freedom of the polarization vector, we can write $v$ and $\bar{v}$ as 
\begin{equation}
v_i=(1,w_i)\;,\quad \bar{v}=(1,\bar{w}_i)\;.    
\end{equation}
When we twist the operators by setting $\bar{w}_i=\bar{z}_i$, the correlator also preserves certain nilponent supercharge. The twisted operators are in its cohomology while the antiholomorphic twisted translations are exact. Therefore, the twisted correlator are meromorphic functions of $z_i$ only.

%% file: sections/Mellinrep.tex
\section{Mellin representation}\label{Sec:Mellin}
It has been commonly advertised that  Mellin space \cite{Mack:2009mi,Penedones:2010ue} is a natural language for discussing holographic correlators. In this formalism, the connected correlators are expressed as a multi-dimensional inverse Mellin transformation
\begin{align}\label{defMellin5pt}
\langle \mathcal{O} (x_1) \dots \mathcal{O} (x_5) \rangle_{\rm conn} = \int [d\delta]  \mathcal{M} (\delta_{ij})\, \prod_{1\leq i<j\leq 5} \Gamma(\delta_{ij})(x_{ij}^2)^{-\delta_{ij}}\;,
\end{align}
 where the Mellin-Mandelstam variables satisfy
 \begin{equation}
 \delta_{ij}=\delta_{ji}\;,\quad \delta_{ii}=-\Delta_i\;,\quad\sum_{j}\delta_{ij}=0\;.
 \label{eq:ConstraintsMellinMand}
 \end{equation}
 The function $\mathcal{M}(\delta_{ij})$ encodes the dynamical information and is referred to as the Mellin amplitude.
Note that this definition is a bit schematic. To be precise, both the correlator and the Mellin amplitude also depend on R-symmetry structures. However, for the moment we will suppress this  dependence to emphasize the analytic structure related to spacetime. One of the reasons that Mellin amplitudes is convenient for describing scattering in AdS is they are meromorphic functions of the Mellin-Mandelstam variables. This follows directly from the existence of the OPE in CFT. Moreover, in the supergravity limit, the poles are associated with the exchanged single-trace particles in AdS. This makes the Mellin amplitudes have similar analytic structure as tree-level scattering amplitudes in flat space and allows us to apply flat-space intuitions in AdS.
 
\begin{figure}[h]
\centering
\includegraphics[scale=0.4]{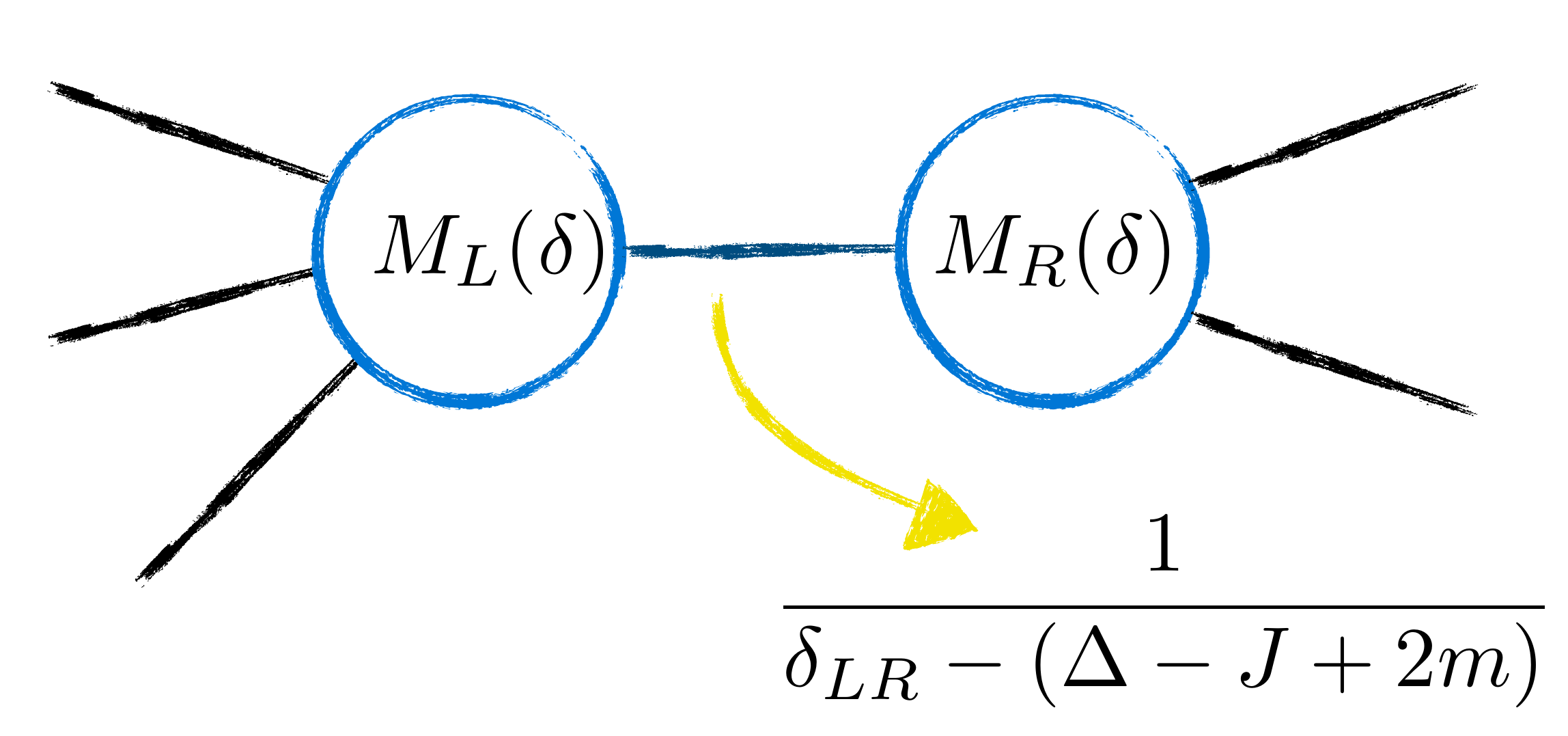}
\caption{ Mellin amplitudes have poles correponding to the exchange of single-trace operators. The residues at the poles are associated with lower-point Mellin amplitudes. In the channel depicted in the figure, we have $n=5$ and $q=3$. The Mellin amplitude on the left has four points while the one on the right has only three. }
    \label{fig_factorization}
\end{figure}

More precisely, the exchange of a conformal primary operator with spin $J$ and dimension $\Delta=\tau+J$ in a channel is represented by a series of poles in the Mellin amplitude, labelled by $m=0,1,2,\ldots$, starting from the conformal twist $\tau$
\begin{align}
\mathcal{M} \approx \frac{\mathcal{Q}_m(\delta_{ij})}{\delta_{LR}-(\tau+2m)}, \ \ \ \ \delta_{LR} = \sum_{a=1}^{q}\sum_{b=q+1}^{n}\delta_{ab}.\label{eq:factorizationequation}
\end{align}
Here, the exchange channel divides the external particles into two sets which we refer to as L and R. We label the particles in L from $1$ to $q$ and the ones in R from $q+1$ to $n$. $\delta_{LR}$ is the Mandelstam variable in this channel. The residues 
$\mathcal{Q}_m(\delta_{ij})$ have nontrivial structures. They are related to the lower-point Mellin amplitudes $\mathcal{M}_L$ and $\mathcal{M}_R$ for the $(q+1)$- and $(n-q+1)$-point functions involving particles in L and R respectively (Figure \ref{fig_factorization}). The extra external state in each lower-point amplitude is the exchanged particle which has now been put on-shell. This is the basic idea of Mellin factorization \cite{Fitzpatrick:2011ia,Goncalves:2014rfa}. In fact, it is very similar to the factorization of amplitudes in flat space which has been studied for a long time. However, there are also important differences. In flat space, the poles are located at the squared masses of the exchanged particles. In Mellin space, as already pointed out, the squared mass is replaced by the conformal twist and there is in general a series of poles for each particle which are labelled by $m$ in  (\ref{eq:factorizationequation}). These are related to the conformal descendants. However, in theories with special spectra such as $AdS_5\times S^5$ IIB supergravity, the series usually truncates. For example, for $p=2$ the series truncates at $m=0$ and contains just one term. Moreover, compared to flat-space amplitudes, the lower-point Mellin amplitudes also appear in the residue $\mathcal{Q}_m$ in a more complicated way. The precise expression for the residues depends on the spin of the operator that is exchanged. The goal of the following subsection is to explain all the details of this formula. In particular, we will present the explicit residue formulas for exchanged fields with spins up to 2. We should emphasize that the structure of factorization for the general $\langle pp222\rangle$ five-point functions will turn out to be far richer than for the simple case of $p=2$ which was analyzed in \cite{Goncalves:2019znr}. In particular, we will see poles with $m\geq 1$. 

Note  that for the five-point function $G_p$ with $p>2$ there are three non-equivalent factorization channels which we choose to be 
\begin{equation}\label{channelsFactorization}
    \begin{split}
    {}& (12):\quad\quad \langle pp \star \rangle  \, \langle \star222 \rangle \;,\\
    {}&(45):\quad\quad \langle 22 \star \rangle \,  \langle \star pp2 \rangle\;,\\ 
    {}&(13):\quad\quad \langle 2p \star\rangle  \, \langle  \star p22\rangle \;,
    \end{split}
\end{equation}
In each of them there are exchanged primary operators with spins ranging from $0$ to $2$ as will be discussed in the following subsection. 

\subsection{Melllin factorization}\label{Subsec:Mellinfactorization}
To discuss Mellin factorization, we need to be more explicit about what fields can be exchanged as they give rise to different lower-point functions. The problem of enumerating exchanged fields reduces to finding all the possible cubic vertices $s_{k_1}s_{k_2}X$ where $s_k$ is the scalar field dual to the superconformal primary $\mathcal{O}_k$ and $X$ is a field to be determined. This problem already appears in the case of four-point functions and therefore the answer is also the same. The possible cubic vertices are determined by two conditions. The first is the R-symmetry selection rule. The second is the condition that the cubic vertices cannot be extremal\footnote{It also follows that four-point functions cannot be extremal or next-to-extremal. In particular, we do not have the four-point functions $\langle 4222\rangle$ and $\langle 6222\rangle$.}. These determine the possible exchange fields to be
 \cite{Rastelli:2016nze,Rastelli:2017udc} 
\begin{eqnarray}\label{possibleexchanges}
\nonumber    &&\{k_1,k_2\}=\{p,p\}\;:\quad X=s_2\;,\;A_{2,\mu}\;,\;\varphi_{2,\mu\nu}\;.\\
 &&\{k_1,k_2\}=\{2,2\}\;:\quad X=s_2\;,\;A_{2,\mu}\;,\;\varphi_{2,\mu\nu}\;,\\
\nonumber       &&\{k_1,k_2\}=\{2,p\}\;:\quad X=s_p\;,\;A_{p,\mu}\;,\;\varphi_{p,\mu\nu}\;.
\end{eqnarray}
Here $s_k$ is a scalar field and is dual to the superconformal primary $\mathcal{O}_k$ which has dimension  $\Delta=k$ and transform in the $[0,k,0]$ representation of $SU(4)$. $A_{k,\mu}$ is a vector field and is dual to a spin-1 operator $\mathcal{J}_{k,\mu}$ which has dimension $\Delta=k+1$ and transforms in the $[1,k-2,1]$ representation. $\varphi_{k,\mu\nu}$ is a spin-2 tensor field and is dual to a spin-2 operator $\mathcal{T}_{k,\mu\nu}$ which has dimension $\Delta=k+2$ and representation $[0,k-2,0]$. When $k=2$, $A_{2,\mu}$ is the graviphoton and $\varphi_{2,\mu\nu}$ is the graviton. Their dual operators are correspondingly the R-symmetry current and the stress energy tensor.  

Let us emphasize again that in this subsection we will only focus on the Mellin-Mandelstam variable dependence. Both $\mathcal{M}_L$ and $\mathcal{M}_R$ in fact also depend on R-symmmtry variables. Therefore in the residues $\mathcal{Q}_m$ there is also a gluing of the lower-point R-symmetry structures. However, this gluing is purely group theoretic. To avoid distracting the reader from the discussion of the dynamics, we will leave the details of R-symmetry gluing to Appendix \ref{Sec:rSymmetryGluing}. Alternatively, we can view the discussion in this subsection as the Mellin factorization for each R-symmetry structure. 

\subsubsection{Exchange of scalars}
The simplest example of factorization is the exchange of a scalar operator with dimension $\Delta$. The resulting $\mathcal{M}_L$ and $\mathcal{M}_R$ are again scalar Mellin amplitudes. Nevertheless, this example contains most of the features we shall need. In particular, the $m$ dependence will be shared in the spinning cases. Therefore, we will first analyze this case in detail.  The residue $\mathcal{Q}_m$ introduced in \eqref{eq:factorizationequation} is given in \cite{Goncalves:2014rfa}
\begin{align}
&\mathcal{Q}_m    =
\frac{-2\Gamma(\Delta) m!}{ 
 \left(1+\Delta-\frac{d}{2}  \right)_m }
 L_m R_m\,,\label{eq:QmScalarDefinition}
\end{align}
where $L_m$ is related to $\mathcal{M}_L$ by\footnote{Notice that $\mathcal{M}_L(\delta_{ab}+n_{ab})$ is well defined when the  Mellin-Mandelstam variables satisfy the pole condition \eqref{eq:factorizationequation}, in addition to their constraints \eqref{eq:ConstraintsMellinMand}. The parallel with scattering amplitudes makes this point clear.} 
\begin{align}
&L_m =\sum_{n_{ab}\ge 0 \atop \sum n_{ab}=m} 
\mathcal{M}_L(\delta_{ab}+n_{ab})  
\prod_{1\le a<b\le q} \frac{ 
\left(\delta_{ab}\right)_{n_{ab}}}{n_{ab}!}\label{eq:Lmdefinition}
\end{align}
and similarly for $R_m$. Notice that here and in the following we will often leave the spacetime dimension $d$ unspecified, but it should always be set to $4$. This equation has several interesting consequences, which will become more evident after analyzing a few examples. Let us start with a three-point Mellin amplitude for $\mathcal{M}_L$, which is just a constant $c$.
In this case, recalling that $\delta_{12}=\tfrac{1}{2} (\Delta_1+\Delta_2-\delta_{LR})$ and $\delta_{LR}$ is set to $\Delta+2m$ by the pole condition \eqref{eq:factorizationequation}, equation \eqref{eq:QmScalarDefinition} with $q=2$ immediately gives
\begin{align}
&\mathcal{M}_L^{\textrm{3-pt}}=c\implies L_m =
c\frac{\big(\bar{\delta}_{LR}\big)_m}{m!}\,,
\qquad
\bar{\delta}_{LR}:=\tfrac{1}{2} (\Delta_1+\Delta_2-\Delta)-m\,.
\label{eq:threeptFactorization}
\end{align}
Factorizing a five-point function leads to a three-point function and a four-point function. For $\langle pp222\rangle$, there are three inequivalent factorization channels, which can be chosen to be $(12)$, $(45)$ and $(13)$. From (\ref{possibleexchanges}), we know that the exchanged scalar operators in these three channels have twists $2$, $2$ and $p$ respectively. Thus, $\bar{\delta}_{LR}$ in each case is given by 
\begin{equation}\label{deltaLR}
    \begin{split}
    {}& (12):\quad\quad \bar{\delta}_{LR} = p-1-m\;,\\
    {}&(45):\quad\quad \bar{\delta}_{LR} = 1-m\;,\\ {}&(13):\quad\quad\bar{\delta}_{LR} = 1-m\;,
    \end{split}
\end{equation}
and the correspoding values of $\delta_{LR}$ are $2+m,2+m,p+m$.
After plugging these values in (\ref{eq:threeptFactorization}), it is straightforward to see that the residue vanishes for $m>0$ in the channels $(13)$ and $(45)$, and for $m\geq p-1$ in the channel $(12)$\footnote{The zeros in these pochhammer symbols are exactly at a position to avoid a double pole, formed by one coming from the explicit Gamma functions in the definition and the other from the factorization formula (\ref{eq:factorizationequation}). }. Naively, one would conclude that in the $(12)$ channel the number of poles increases with $p$.   However, this is too fast since the other part $R_m$ can give more constraints. To see this explicitly, let us look at a  four-point Mellin amplitude which has the following generic form
\begin{align}
&\mathcal{M}_R^{\textrm{4pt}}=\frac{c_1\delta_{45}^2+c_2\delta_{45}+c_3}{\delta_{34}-1}+c_4+c_{5}\delta_{34}+c_{6}\delta_{45}\implies R_m = \frac{1}{m! }\bigg[\frac{c_1 m \delta _{45} \left(\delta _{45}+1\right) (3 - m)_{m-1}}{\delta_{34}-1}\nonumber\\
& +\left(\frac{c_1 \delta _{45}^2+c_2 \delta _{45}}{\delta_{34}-1}+c_4 \right) (2 - m)_m+\frac{c_3 (1 - m)_m}{\delta_{34}-1}+(c_5\delta_{34}+c_6\delta_{45})(3-m)_m\bigg]\;.\label{eq:prototypical4Mellin}
\end{align}
Here we have evaluated the expression  at the pole $\delta_{LR}=\tau+2m$.  It follows that $R_m$ vanishes for this four-point Mellin $\mathcal{M}_R$ for $m\ge 3$ and therefore the number of poles does not increase for arbitrary value of $p$. Let us also emphasize that all four-point Mellin amplitudes that appear in the OPE of the correlator $\langle pp222\rangle$ have this structure as can be checked in Appendix \ref{Sec:higherkkSmult}.

Let us note that the absence of poles for $m\geq p-1$ can also be understood from the pole structure of the Mellin integrand. The Gamma functions in the definition of Mellin amplitude already have poles in this location and a pole in the Mellin amplitude at $m\geq p-1$ would give rise to a double pole. Such double poles are associated with the appearance of anomalous dimension \cite{Penedones:2010ue,Rastelli:2016nze,Rastelli:2017udc}, which we do not expect at this order. On the other hand, at the moment we do not have a direct physical argument for the truncation of poles at $m \ge 3$. Finally, this truncation continues to hold for the factorization formulas when the exchanged operators have spins. This will be analyzed in the following subsubsection. 

\subsubsection{Exchange of operators with spins 1 and 2}
In this subsection we will be interested in studying the contribution of operators with spins. As it turns out, the analysis of the scalar case straightforwardly generalizes to the spinning case. It is convenient to get rid of the Lorentz indices of these operators by contracting them with null polarization vectors 
\begin{align}
\mathcal{O}(x,z) =  \mathcal{O}^{a_1\dots  a_J}(x) z^{a_1}\dots z^{a_J}\;,
\end{align}
where $z^2=0$ ensures the operator is traceless (we refer the reader to Section $3$ of \cite{Goncalves:2014rfa} for a more detailed review). The definition of Mellin amplitudes of one spinning operator and $n$ scalar operators is given by \cite{Goncalves:2014rfa} 
\begin{align}\label{scalardeltaLR}
\langle \mathcal{O}(x_0,z_0) \dots \mathcal{O}_n   \rangle =&\sum_{a_1,\dots,a_J=1}^{n} \prod_{i=1}^{J}(z_0\cdot x_{a_i0})
\int [d\delta]  \mathcal{M}^{\{ a\} }(\delta_{ij}) \prod_{i=1}^n \frac{\Gamma(\delta_i+\{a \}_i )}{(x_{i0}^2) ^{\delta_i+\{a \}_i } }  \prod_{1\leq i<j\leq n} \frac{\Gamma(\delta_{ij})}{(x_{ij}^2)^{\delta_{ij}}}, 
\end{align}
where 
\begin{equation}
    \{a \}_i = \textrm{{\fancy{$\delta$}}}_{i}^{a_1}+\dots+\textrm{{\fancy{$\delta$}}}_{i}^{a_J}, \ \  \delta_{i}= -\sum_{j=1}^n\delta_{ij} ,\, \ \ \ \sum_{i,j=1}^n\delta_{ij} =J-\Delta_0\;.
\end{equation}
We have used $\textrm{{\fancy{$\delta$}}}$ to denote the Kronecker delta so that it can be distinguished from the Mellin-Mandelstam variables $\delta$. 
The Mellin amplitudes $\mathcal{M}^{\{ a\} }$ satisfy certain linear relations that follows from the conformal invariance of the correlator, see equation (46) in \cite{Goncalves:2014rfa}.
Let us first focus on the spinning generalization of (\ref{eq:QmScalarDefinition}) for the conserved currents which reside in the $k=2$ supermultiplet. For exchanging the graviphoton, the residues are given by\footnote{As above we write $d$ to denote the dimension of space-time and we will always set $d=4$.}
\begin{align}
&\mathcal{Q}_m    =
\frac{(d-1)\Gamma(d-2) m!}{ 
 \left(\frac{d}{2}  \right)_m }
\sum_{a=1}^{q}\sum_{b=q+1}^n \delta_{ab} L_m^{a} R_m^{b}\;,
\end{align}
For exchanging the graviton, the residues are
\begin{equation}
    \mathcal{Q}_m = \frac{-(d+1)\Gamma(d-1)m!}{2\left(\frac{d}{2} +1\right)_m}\bigg[\mathcal{Q}_{m}^{(1)} -\left(\frac{1}{2m}+\frac{1}{d} \right)\tilde{L}_m \tilde{R}_m \bigg]\;,
\end{equation}
where
\begin{align}
&\mathcal{Q}_{m}^{(1)}  = \sum_{a,b=1}^{q}\sum_{i,j=q+1}^{n}\delta_{ai}(\delta_{bj}+\textrm{{\fancy{$\delta$}}}
_{b}^{a}\textrm{{\fancy{$\delta$}}}_{j}^{i} )L_m^{ab}R_{m}^{ij},\,\\
&\tilde{L}_m = \sum_{a,b=1}^{q}\delta_{ab} [ L_{m-1}^{ab} ]^{ab}\;,\quad \tilde{R}_m = \sum_{a,b=1}^{q}\delta_{ab} [ R_{m-1}^{ab} ]^{ab}\;.  \nonumber
\end{align}
Here we used the notation $[f(\delta_{ij})]^{ab} = f(\delta_{ij}+\textrm{{\fancy{$\delta$}}}_{i}^{a}\textrm{{\fancy{$\delta$}}}_{j}^{b}+\textrm{{\fancy{$\delta$}}}_{j}^{a}\textrm{{\fancy{$\delta$}}}_{i}^{b})$. The functions $L_m^{a}$ and $L_{m}^{ab}$ (and analogously $R_m^{a}$, $R_m^{ab}$) are defined in the same way as in (\ref{eq:Lmdefinition}). Let us also add that for $m=0$ the second term in $Q_m$ for spin $2$ is zero since both $\tilde{L}_0$ and $\tilde{R}_0$ vanish from the definition. Therefore, the appearance of the pole in $m$ does not lead to a divergence.

These residue formulas for spinning operators clearly are not the full story as there are also non-conserved currents in the multiplets with $k>2$. However, from (\ref{possibleexchanges}) we can see that such non-conserved currents only appear in the channel with $s_2$ and $s_p$. Similar to the scalar case (\ref{deltaLR}), the analysis of the three-point functions requires the truncation at $m=0$. The residues are
\begin{align}
Q_{0} &= 
-\Delta\Gamma(\Delta-1) \sum_{a=1}^{q}\sum_{b=q+1}^n \delta_{ab} L_0^{a} R_0^{b}, & \quad \textrm{for spin $1$}\;, \\
Q_{0} &= 
-\frac{(\Delta+1)\Gamma(\Delta-1)}{2}\sum_{a,b=1}^q\sum_{i,j=q+1}^{n}\delta_{ai}(\delta_{bj}+\delta_{b}^{a}\delta_{j}^{i} )L_0^{ab}R_{0}^{ij}                 & \quad \textrm{for spin $2$}\;.
\end{align}
The most general expressions for factorization with arbitrary external and internal dimensions and $m$ can be found in \cite{Goncalves:2014rfa}. But they are not needed in this paper.

As in the scalar case, the truncation of poles also relies on the form of the spinning four-point amplitudes. They are given in Appendix \ref{Sec:strongcouplingcorr} (see (\ref{eq:currentMellinexample}) and (\ref{eq:StressMellinexample}) for explicit expressions). In particular, they have the same analytic structure as the scalar four-point amplitude (\ref{eq:prototypical4Mellin}) except that now they carry additional indices. As a result, the truncation of poles also holds for the exchange of spinning operators. More precisely, we have the same pole locations as in (\ref{deltaLR}) where the allowed values for $m$ are $m=0,1,2$ for $(12)$ and $m=0$ for $(45)$, $(13)$.

To summarize, the Mellin factorization formulas allow us to reconstruct all the polar part of the amplitude from the lower-point Mellin amplitudes. Furthermore, the spectrum of the theory gives rise to a further simplification where the poles truncate to a finite range independent of $p$.

\subsection{Drukker-Plefka twist in Mellin space}\label{subsec:DrukkerPlefkaTwist}
As we reviewed in the introduction, the two superconformal constraints, namely the chiral algebra and the Drukker-Plefka twist, were both formulated and implemented in position space \cite{Goncalves:2019znr}. To have a more streamlined algorithm, we would like to perform the bootstrap entirely within Mellin space and therefore need to translate such position space constraints into Mellin space. Let us first define the Mellin amplitude more precisely by restoring the R-symmetry dependence suppressed in the definition (\ref{defMellin5pt}). For the $\langle pp222\rangle$ correlator, we have
  \begin{align}\label{defMellinRsymm}
&G_p(x_i,t_i) = \int [d\delta]  \mathcal{M} (\delta_{ij},t_{ij})\, \prod_{1\leq i<j\leq 5} \Gamma(\delta_{ij})(x_{ij}^2)^{-\delta_{ij}}\;,
\end{align}
where $\mathcal{M} (\delta_{ij},t_{ij})$ is a linear combination of the 29 R-symmetry structures listed in (\ref{indepRstructures}).
Usually the implementation of the twists in Mellin space is achieved by using the observation that $x_{ij}^2$ monomials multiplying the Mellin transform (\ref{defMellinRsymm}) can be absorbed into the definition by shifting the Mellin-Mandelstam variables. This gives rise to difference equations in Mellin space. This strategy has been used, for example, in \cite{Zhou:2017zaw,Zhou:2018ofp} to rewrite the superconformal Ward identities in Mellin space for four-point functions. In our case, there are extra complexities.

The issue is that the chiral algebra constraint requires all the operators to be on a two dimensional plane. When the number of operators $n>4$, this cannot be achieved by a conformal transformation and there are relations among the cross ratios.\footnote{In two dimensions, the number of independent cross ratios is $2n-6$ for $n\geq 2$. However, in high enough spacetime dimensions, the number of independent cross ratios is $\frac{n(n-3)}{2}$. The relation for the cross ratios can be written in form of ${\rm det}M=0$ where the matrix $M$ has elements $M_{ij}=x_{ij}^2$. } The meromorphy of the correlator after the chiral algebra twist depends crucially on these relations. On the other hand, these relations do not hold in the definition of the Mellin ampllitude where the locations of the operators are assumed to be general. Therefore, the position space chiral algebra condition cannot be translated into Mellin space using the same strategy. 

By contrast, the Drukker-Plefka twist only imposes conditions on the R-symmetry polarizations and has no restriction on the operator insertions. Therefore, we can use the same trick to implement the Drukker-Plefka twist in Mellin space. More precisely, we can extract a kinematic factor and rewrite (\ref{defMellinRsymm}) in terms of cross ratios (\ref{eq:crossratiosPos}), (\ref{eq:crossratiosRCharge})
\begin{equation}
G_p(x_i,t_i) =  K_p \int d\delta_{ij} \mathcal{M} (\delta_{ij},\sigma_i) \Gamma_{\textrm {pp222}} \,u_{1}^{p-\delta_{12}}u_{2}^{-\delta_{23}}\, u_{3}^{2-\delta_{34}}\, u_{4}^{-\delta_{45}}\, u_{5}^{1-\delta_{15}} \;.
\end{equation}
Here $K_p$ is a kinematic factor  
\begin{equation}
    K_p=\frac{x_{13}^2t_{12}^pt_{34}^2 t_{15} t_{35}}{(x_{12}^2)^{p} (x_{34}^2)^{2}(x_{15}^2x_{35}^2) t_{13}}\;,
\end{equation}
and
\begin{align}
&\Gamma_{\textrm {pp222}}  = \Gamma \left(\delta _{12}\right) \Gamma \left(\delta _{15}\right) \Gamma \left(\delta _{23}\right) \Gamma \left(\delta _{15}-\delta _{23}-\delta _{34}+1\right) \Gamma \left(\delta _{34}\right) \Gamma \left(\delta _{23}+1-\delta _{15}-\delta _{45}\right) \Gamma \left(\delta _{45}\right)\nonumber\\
& \Gamma \left(p-\delta _{12}-\delta _{15}+\gamma _{34}-1\right) \Gamma \left(\delta _{12}-p-\delta _{34}-\delta _{45}+3\right) \Gamma \left(p-\delta _{12}-\delta _{23}+\delta _{45}-1\right)\;.
\end{align}
Moreover, we have chosen $\delta_{12},\delta_{23},\delta_{34},\delta_{45}$ and $\delta_{15}$ as the independent Mellin variables. Performing the Drukker--Plefka twist amounts to setting $t_{ij}\rightarrow x_{ij}^2$, or equivalently $\sigma \rightarrow u$ for the cross ratios. To implement this in practice, we notice that doing the twist reduces to multiplying the Mellin representation of different terms of the correlator $K_p^{-1}G_p(x_i,t_i)$ by monomials  $u_1^{n_1}\,u_2^{n_2}u_3^{n_3}u_4^{n_4}u_5^{n_5}$
\begin{equation}
\mathcal{M}(\delta_{ij},\sigma_i)=\sum_{\{n_i\}}\sigma_1^{n_1}\,\sigma_2^{n_2}\sigma_3^{n_3}\sigma_4^{n_4}\sigma_5^{n_5}\mathcal{M}_{\{n_i\}}(\delta_{ij})\to \sum_{\{n_i\}}u_1^{n_1}\,u_2^{n_2}u_3^{n_3}u_4^{n_4}u_5^{n_5}\mathcal{M}_{\{n_i\}}(\delta_{ij})\;.
\end{equation}
We can absorb them by shifting $\delta_{ij}$ and this has the effect on the Mellin amplitudes by acting with a difference operator 
\begin{equation}
u_1^{n_1}\,u_2^{n_2}u_3^{n_3}u_4^{n_4}u_5^{n_5}\mathcal{M}_{\{n_i\}}(\delta_{ij}) \rightarrow  \mathbb{D}_{n_1,\ldots,n_5}\circ \mathcal{M}_{\{n_i\}}(\delta_{ij})\;,   
\end{equation}
where the explicit action of $\mathbb{D}_{n_1,\ldots,n_5}$ reads
\begin{align}
&\mathbb{D}_{n_1,\ldots,n_5}\circ \mathcal{M}_{\{n_i\}}(\delta_{ij})    =  \mathcal{M}_{\{n_i\}}(\delta_{12} +n_1,\delta_{23}+n_2,\dots ) \times \left(\delta _{12}\right)_{n_1} \left(\delta _{15}\right)_{n_5} \left(\delta _{23}\right)_{n_2} \left(\delta _{34}\right)_{n_3} \left(\delta _{45}\right)_{n_4}\nonumber\\
&\left(\delta _{15}-\delta _{23}-\delta _{34}+1\right)_{n_5-n_2-n_3} \left(\delta _{23}-\delta _{15}-\delta _{45}+1\right)_{n_2-n_4-n_5}  \left(p-\delta _{12}-\delta _{15}+\delta _{34}-1\right)_{n_3-n_1-n_5} \nonumber\\
&\left(\delta _{12}-p-\delta _{34}-\delta _{45}+3\right)_{n_1-n_3-n_4} \left(p-\delta _{12}-\delta _{23}+\delta _{45}-1\right)_{n_4-n_1-n_2}\;.
\end{align}
The various Pochhammer symbols come from comparing the shifted Gamma factor with the one in the Mellin representation definition. The full difference operator from the Drukker-Plefka twist, denoted as $\mathbb{D}_{\rm DP}$, is then a sum of such operators acting on different R-symmetry structures. As we explained in Sec. \ref{Subsec:DrukkerPlefkatwist}, the twisted correlator is just a constant in position space. Following \cite{Rastelli:2016nze,Rastelli:2017udc}, we should interpret its Mellin amplitude as zero. Therefore, the Drukker-Plefka twist condition becomes in Mellin space
\begin{equation}
\mathbb{D}_{\rm DP}\circ \mathcal{M}(\delta_{ij},\sigma_i)=0\;,\label{eq:DrukkerPlefkadifferenceOPE}
\end{equation}
which explicitly reads
\begin{equation}
\sum_{\{n_i\}}\mathbb{D}_{n_1,\ldots,n_5}\circ \mathcal{M}_{\{n_i\}}(\delta_{ij})=0\;.
\end{equation}
The implications of this equation are discussed in the following section.

%% file: sections/bootstrapMellin.tex
\section{Bootstrapping five-point Mellin amplitudes}\label{Sec:bootstrapMellin}
\subsection{Strategy and ansatz}

After introducing all the necessary ingredients, we are now ready to state our strategy. Our strategy is comprised of three steps. First, we start by formulating an ansatz in Mellin space which is based on our analysis of the analytic structure of the Mellin amplitudes. Second, we impose the Mellin factorization condition which is the statement that the pole residues should be correctly reproduced by the lower-point amplitudes. Finally, we implement the Drukker-Plefka twist in Mellin space and completely fix the ansatz. In the following, we explain the details of each step.

\vspace{0.3cm}
\noindent{\it Step 1: Ansatz}
\vspace{0.2cm}

\noindent As we emphasized in the previous section, Mellin amplitudes are merophormic functions with simple poles corresponding to exchanging single-trace operators and residues related to lower-point amplitudes via factorization. Based 
 on this, we have the following ansatz for the $\langle pp222\rangle$ Mellin amplitude 
\begin{align}
\nonumber \mathcal{M}(\delta_{ij},t_{ij}) =& \sum_{m=0}^{2} \frac{A_{m}(\delta_{ij},t_{ij})}{(\delta_{12} +1+m-p)}+\sum_{\bar{a}=1,2,a=3,4,5} \frac{B_{\bar{a}a}(\delta_{ij},t_{ij})}{\delta_{\bar{a}a}-1}+ \sum_{3\leq a < b \leq5} \frac{C_{ab}(\delta_{ij},t_{ij})}{\delta_{ab}-1}\\
&+D(\delta_{ij},t_{ij})\;.\label{eq:ansatzMellin}
\end{align}
Here $A_{m}(\delta_{ij},t_{ij})$ is a rational function with possible poles in $\delta_{34}$, $\delta_{35}$, $\delta_{45}$. In particular, it includes simultaneous poles which correspond to double exchange processes in the (12), (34) channels {\it etc}. Similarly, $B_{\bar{a}a}(\delta_{ij},t_{ij})$ is a rational function with possible poles in $\delta_{kl}$ at $\delta_{kl}=1$. 
 The labels $k$, $l$ need to satisfy $k,l \neq \bar{a},a$  but can be both from the set $\{3,4,5\}$, or belong to different sets $\{1,2\}$ and $\{3,4,5\}$, see equation \eqref{othertermsinAnsatz}. To avoid double counting, $C_{jk}(\delta_{ij},t_{ij})$ and $D(\delta_{ij},t_{ij})$ do not have poles and they
are polynomial functions of the Mellin-Mandelstam variables.
Note that here we have also used our Mellin factorization analysis for the subleading poles from Section \ref{Subsec:Mellinfactorization}.  We imposed that the poles in the (12) channel truncate to $m=0,1,2$.  

More concretely, the function  $A_m(\delta_{ij},t_{ij}) $ in the ansatz has the following form
\begin{align}
\label{eq:P12form}
A_{m} (\delta_{ij},t_{ij}) = \frac{A_{34,m}(\delta_{ij},t_{ij})}{\delta_{34}-1}+\frac{A_{35,m}(\delta_{ij},t_{ij})}{\delta_{35}-1}+\frac{A_{45,m}(\delta_{ij},t_{ij})}{\delta_{45}-1}+ A_{\emptyset,m}(\delta_{ij},t_{ij})\;,
\end{align}
where $A_{34,m}$, $A_{35,m} $ and $A_{45,m}$ are polynomials of degree $2$ and $A_{\emptyset,m}$ is a polynomial of degree $1$. Written explicitly, $A_{34,m}$ reads
\begin{align}
A_{34,m}(\delta_{ij},t_{ij}) = \sum_{\alpha_i}^{\alpha_1+\alpha_2+\alpha_3\leq 2} \sum_{I=1}^{29} a^I_{34,m,\{\alpha_i\}} \delta_{23}^{\alpha_1}\delta_{25}^{\alpha_2}\delta_{45}^{\alpha_3} \mathcal{T}_I,
\end{align}
where $\{\delta_{23},\delta_{25},\delta_{45}\}$ are chosen to be the independent Mellin-Mandelstam variables in addition to $\delta_{12}$ and $\delta_{34}$ which already appear in the poles. We have also used $\{\mathcal{T}_I\}$ to denote collectively the 29 independent R-symmetry structures in (\ref{indepRstructures}). The expressions for $A_{35,m}$, $A_{45,m}$ are similar. The polynomial $A_{\emptyset,m}$ is given by 
\begin{align}
A_{\emptyset,m} = \sum_{\alpha_i}^{\alpha_1+\alpha_2+\alpha_3+\alpha_4\leq 1} \sum_{I=1}^{29} a^I_{\emptyset,m,\{\alpha_i\}} \delta_{23}^{\alpha_1}\delta_{25}^{\alpha_2}\delta_{45}^{\alpha_3}\delta_{34}^{\alpha_4} \mathcal{T}_I.
\end{align}
The other terms in the ansatz are similar and are given by
\begin{align}
\nonumber B_{13} =& \sum_{I=1}^{29}\bigg(\sum_{\alpha_i}^{\alpha_1+\dots\alpha_3\leq 2}\bigg[\frac{ b^I_{13,24,\{\alpha_i\}} \delta_{15}^{\alpha_1} \delta_{23}^{\alpha_2}\delta_{45}^{\alpha_3} }{\delta_{24}-1} +\frac{b^I_{13,45,\{\alpha_i\}}\delta_{15}^{\alpha_1} \delta_{23}^{\alpha_2}\delta_{34}^{\alpha_3}}{\delta_{45}-1} +\frac{b^I_{13,25,\{\alpha_i\}}\delta_{23}^{\alpha_1} \delta_{34}^{\alpha_2}\delta_{45}^{\alpha_3}}{\delta_{25}-1}\bigg] 
\label{othertermsinAnsatz}
\\
&+ \sum_{\alpha_i}^{\alpha_1+\dots\alpha_4\leq 1} b^I_{13,\emptyset,\{\alpha_i\}}\delta_{23}^{\alpha_1}\delta_{34}^{\alpha_2} \delta_{45}^{\alpha_3}\delta_{15}^{\alpha_4}\bigg)\mathcal{T}_I\;,\\
\nonumber C_{34} =& \sum_{\alpha_i}^{\alpha_1+\dots\alpha_3\leq 1}\sum_{I=1}^{29} c^I_{34,\{\alpha_i\}} \delta_{12}^{\alpha_1}\delta_{23}^{\alpha_2}\delta_{45}^{\alpha_3}\delta_{15}^{\alpha_4}\mathcal{T}_I\;,\\
\nonumber D =& \sum_{I=1}^{29} d^I \mathcal{T}_I \;.
\end{align}
In making the ansatz we have assumed that the degrees of various polynomials are the same as in the $p=2$ correlator. This is expected from the flat-space limit which is related to the high energy limit of the Mellin amplitude \cite{Penedones:2010ue}. This can also be confirmed by Mellin factorization, which will be used in greater detail in the next step.\footnote{For example, it is straightforward to see that these are the correct degrees when exchanging scalar operators. Exchanging vector or tensor fields is a bit more nontrivial but it is possible to check that the degrees are correct. The only subtle point which avoids the factorization argument is the degree of the regular piece. However, it is natural to assume that the degree is the same as the $p=2$ case so that it has the same high energy growth as the other terms.}

\vspace{0.3cm}
\noindent{\it Step 2: Mellin factorization}
\vspace{0.2cm}

The second step of our strategy is  to impose Mellin factorization. 
As explained in the previous section, all the polar terms of the Mellin amplitude can be completely fixed  in terms of the lower-point Mellin amplitudes. For the $\langle pp222\rangle$ five-point function, all these lower-point amplitudes are known and are given in Appendix \ref{Sec:strongcouplingcorr}. These lower point functions depend on $R$-symmetry polarization vectors. One important detail which we did not discuss is how to glue together the $R$-symmetry structures in the lower-point functions using the representation of the exchanged fields. This step is explained in detail in Appendix \ref{Sec:rSymmetryGluing}. Thus all terms in the ansatz (\ref{eq:ansatzMellin}), except for the regular term $D$,  can be fixed by using this factorization procedure. Note that the number of coefficients that remain unfixed in the ansatz is quite low as $D$ is just a constant with respect to the Mellin-Mandelstam variables. It can depend only on the linear combination coefficients of the 29 $R$-symmetry structures. 

\vspace{0.3cm}
\noindent{\it Step 3: Drukker-Plefka twist}
\vspace{0.2cm}

The final step is to impose the Drukker-Plefka twist. As explained in Section \ref{subsec:DrukkerPlefkaTwist} this twist can be phrased in terms of a difference operator $\mathbb{D}_{\rm DP}$ acting on the Mellin amplitude, see (\ref{eq:DrukkerPlefkadifferenceOPE}).  This relates the regular part with the singular part already fixed by factorization and completely fixes the remaining coefficients\footnote{At the same time the Drukker-Plefka twist provides a very non trivial consistency check for the procedure of extracting correlation functions of super-descendants and gluing of R-symmetry structures described in Appendix.}. 

Using this strategy, we obtain the $\langle pp222\rangle$ Mellin amplitudes in a closed form for arbitrary $p$. The final result for the Mellin amplitudes will be presented in the next section\footnote{It would also be interesting to extend this analysis to the first correction in $\alpha'$. One promising candidate is the $p=2$ case since it is more symmetric and we can also use the known results for the four-point function as an input \cite{Goncalves:2014ffa}. }. 

\subsection{Mellin amplitude for $p=2$}
Due to the many R-symmetry structures involved, the expression for the full Mellin amplitude appears to be quite complicated at first sight. Therefore, before we present the Mellin amplitude for general $p$, let us first revisit the $p=2$ result of \cite{Goncalves:2019znr} and present it in a simpler way. 

When $p=2$, the amplitude is symmetric under permutations of all the five external points. The 22 R-symmetry structures also split into two classes and within each class the structures are related by permutations. The first class is the pentagon contraction 
\begin{equation}
P_a=\{t_{12}t_{23}t_{34}t_{45}t_{15},\ldots\}\;,\quad a=1,2,\ldots, 12\;,
\end{equation}
which includes $P^{({\rm I,II})}_{a_3a_4a_5}$ in (\ref{indepRstructures}). The second class is the contraction of three points times the contraction of the remaining two points
\begin{equation}
T_a=\{t_{12}t_{23}t_{13}t_{45}^2,\ldots\}\;,\quad a=1,2,\ldots, 10\;,
\end{equation}
which includes $T^{({\rm I,II,III})}_{a_3a_4a_5}$ in (\ref{indepRstructures}). The full amplitude can be written as 
\begin{equation}
\mathcal{M}_{p=2}=\sum_{a=1}^{12} \mathcal{M}^P_a P_a+\sum_{a=1}^{10} \mathcal{M}^T_a T_a\;.
\end{equation}
It is sufficient to determine the coefficient amplitudes $\mathcal{M}^P_1$ and  $\mathcal{M}^T_1$ as the rest can be obtained by permutations. We find 
\begin{equation}\label{M1p}
\begin{split}
 \mathcal{M}^P_1={}&4\sqrt{2} \bigg\{\frac{(\delta_{14}+\delta_{24})(\delta_{13}+\delta_{14})}{(\delta_{12}-1)(\delta_{34}-1)}+\frac{(\delta_{14}+\delta_{24})(\delta_{24}+\delta_{25})}{(\delta_{12}-1)(\delta_{45}-1)}+\frac{(\delta_{25}+\delta_{35})(\delta_{24}+\delta_{25})}{(\delta_{23}-1)(\delta_{45}-1)}\\
{}& +\frac{(\delta_{25}+\delta_{35})(\delta_{13}+\delta_{35})}{(\delta_{23}-1)(\delta_{15}-1)}+\frac{(\delta_{13}+\delta_{35})(\delta_{13}+\delta_{14})}{(\delta_{15}-1)(\delta_{34}-1)}+\frac{1}{2}\bigg(\frac{\delta_{35}}{\delta_{12}-1}+\frac{\delta_{14}}{\delta_{23}-1}\\
{}&+\frac{\delta_{25}}{\delta_{34}-1}+\frac{\delta_{13}}{\delta_{45}-1}+\frac{\delta_{24}}{\delta_{15}-1}\bigg)-2\bigg\}\;,
\end{split}
\end{equation}
\begin{equation}\label{M1T}
\begin{split}
 \mathcal{M}^T_1={}&-2\sqrt{2}\bigg(\frac{(\delta_{13}+\delta_{14})(\delta_{23}+\delta_{24})}{(\delta_{12}-1)(\delta_{34}-1)}+\frac{(\delta_{13}+\delta_{15})(\delta_{23}+\delta_{25})}{(\delta_{12}-1)(\delta_{35}-1)}+\frac{(\delta_{14}+\delta_{15})(\delta_{24}+\delta_{25})}{(\delta_{12}-1)(\delta_{45}-1)}\bigg)\;\nonumber.
\end{split}
\end{equation}
It is clear that terms of the same structure are related by the permutations preserved by the R-symmetry structure. We will see that the Mellin amplitude for general p also has similar structures. 
\subsection{Mellin amplitudes for general $p$}
For $p>2$, we no longer have the full permutation symmetry and there are seven types of R-symmetry structures as we discussed in Section \ref{Subsec:Rsymm}. The Mellin amplitude can be written as a sum over all the inequivalent R-symmetry structures
\begin{align}\label{M5ptgeneralp}
\mathcal{M}_p={}&\sum_{\mathcal{I}_1}\mathcal{M}^{P, ({\rm I})}_{a_3a_4a_5} P^{({\rm I})}_{a_3a_4a_5}+\sum_{\mathcal{I}_1}\mathcal{M}^{P, ({\rm II})}_{a_3a_4a_5} P^{({\rm II})}_{a_3a_4a_5}+\mathcal{M}^{T,({\rm I})}_{345} T^{({\rm I})}_{345}+\sum_{\mathcal{I}_2}\mathcal{M}^{T, ({\rm II})}_{a_3a_4a_5} T^{({\rm II})}_{a_3a_4a_5}\nonumber\\
{}&+\sum_{\mathcal{I}_3}\mathcal{M}^{T, ({\rm III})}_{a_1a_2a_3a_4a_5} T^{({\rm III})}_{a_1a_2a_3a_4a_5}+\mathcal{M}^{N,({\rm I})}_{345} N^{({\rm I})}_{345}+\sum_{\mathcal{I}_1}\mathcal{M}^{N, ({\rm II})}_{a_3a_4a_5} N^{({\rm II})}_{a_3a_4a_5}\;,
\end{align}
where the sets $\mathcal{I}_{1,2,3}$ contain the following permutations
\begin{align}
\mathcal{I}_1={}&\{(3,4,5),(3,5,4),(4,3,5),(4,5,3),(5,3,4),(5,4,3)\}\;,\nonumber\\
\mathcal{I}_2={}&\{(3,4,5),(4,3,5),(5,3,4)\}\;,\\
\mathcal{I}_3={}&\{(1,2,3,4,5),(1,2,4,3,5),(1,2,5,3,4),(2,1,3,4,5),(2,1,4,3,5),(2,1,5,3,4)\}\nonumber
\end{align}
The coefficient Mellin amplitudes are given as follows. For the structures of $P^{({\rm I})}_{a_3a_4a_5}$, $P^{({\rm II})}_{a_3a_4a_5}$, the coefficients are
\begin{align}
\mathcal{M}^{P, ({\rm I})}_{a_3a_4a_5}={}&2\sqrt{2}p\bigg\{\frac{2}{p}\frac{\delta_{1a_4}+\delta_{2a_4}}{\delta_{12}-p+1}\bigg(\frac{\delta_{1a_3}+\delta_{1a_4}}{\delta_{a_3a_4}-1}+\frac{\delta_{2a_4}+\delta_{2a_5}}{\delta_{a_4a_5}-1}\bigg)+\frac{1}{p}\frac{\delta_{a_3a_5}}{\delta_{12}-p+1}\nonumber\\
{}&+\frac{p-2}{p}\frac{\delta_{1a_4}+\delta_{2a_4}-1}{\delta_{12}-p+2}\bigg(\frac{\delta_{1a_3}+\delta_{1a_4}}{\delta_{a_3a_4}-1}+\frac{\delta_{2a_4}+\delta_{2a_5}}{\delta_{a_4a_5}-1}-1\bigg)\nonumber\\
{}&-\frac{(p-2)(p-3)}{2p}\frac{\delta_{a_3a_5}}{\delta_{12}-p+3}+\frac{(\delta_{2a_5}+\delta_{a_3a_5})(\delta_{2a_4}+\delta_{2a_5})}{(\delta_{2a_3}-1)(\delta_{a_4a_5}-1)}+\frac{(\delta_{1a_3}+\delta_{a_3a_5})(\delta_{1a_3}+\delta_{1a_4})}{(\delta_{1a_5}-1)(\delta_{a_3a_4}-1)}\nonumber\\
{}&+\frac{p}{2}\frac{(\delta_{1a_3}+\delta_{a_3a_5})(\delta_{2a_5}+\delta_{a_3a_5})}{(\delta_{1a_5}-1)(\delta_{2a_3}-1)}+\frac{1}{2}\bigg(\frac{\delta_{2a_4}}{\delta_{1a_5}-1}+\frac{\delta_{1a_4}}{\delta_{2a_3}-1}\bigg)\nonumber\\
{}&+\frac{p-1}{p}\bigg(\frac{\delta_{2a_5}}{\delta_{a_3a_4}-1}+\frac{\delta_{1a_3}}{\delta_{a_4a_5}-1}\bigg)+\frac{6-7p}{2p}\bigg\}\;,
\end{align}
\begin{align}
\mathcal{M}^{P, ({\rm II})}_{a_3a_4a_5}={}&\sqrt{2}p^2\bigg\{\frac{(\delta_{12}+\delta_{2a_5})(\delta_{2a_5}+\delta_{a_3a_5})}{(\delta_{1a_5}-1)(\delta_{2a_3}-1)}+\frac{(\delta_{12}+\delta_{1a_4})(\delta_{1a_4}+\delta_{a_3a_4})}{(\delta_{2a_4}-1)(\delta_{1a_3}-1)}\nonumber\\
{}&+\frac{(\delta_{12}+\delta_{2a_5})(\delta_{12}+\delta_{1a_4})}{(\delta_{1a_5}-1)(\delta_{2a_4}-1)}-\frac{(p-2)\delta_{12}}{(\delta_{1a_5-1})(\delta_{2a_4}-1)}+\frac{2\delta_{12}}{p(\delta_{a_4a_5}-1)}\bigg(\frac{\delta_{1a_4}}{\delta_{2a_3}-1}+\frac{\delta_{2a_5}}{\delta_{1a_3}-1}\bigg)\nonumber\\
{}&-\frac{2(p+1)\delta_{12}}{p^2(\delta_{a_4a_5}-1)}+\frac{(p-2)(\delta_{12}+1)+\delta_{a_3a_4}}{p(\delta_{1a_5}-1)}+\frac{(p-2)(\delta_{12}+1)+\delta_{a_3a_5}}{p(\delta_{2a_4}-1)}\nonumber\\
{}&+\frac{1-p}{p}\bigg(\frac{\delta_{1a_4}}{\delta_{2a_3}-1}+\frac{\delta_{2a_5}}{\delta_{1a_3}-1}\bigg)+\frac{p-2}{p}\bigg\}\;.
\end{align}
Upon setting $p=2$, the two coefficient amplitudes become degenerate up to permutations and reproduce $\mathcal{M}_1^P$ in (\ref{M1p}). The coefficient Mellin amplitudes of $T^{({\rm I})}_{345}$,  $T^{({\rm II})}_{a_3a_4a_5}$ and $T^{({\rm III})}_{a_1a_2a_3a_4a_5}$ are given by
\begin{equation}
\begin{split}
\mathcal{M}^{T, ({\rm I})}_{345}={}&-2\sqrt{2}\bigg\{\frac{1}{\delta_{12}-p+1}\bigg(\frac{(\delta_{1a_3}+\delta_{1a_4})(\delta_{2a_3}+\delta_{2a_4})}{\delta_{a_3a_4}-1}+\frac{(\delta_{1a_3}+\delta_{1a_5})(\delta_{2a_3}+\delta_{2a_5})}{\delta_{a_3a_5}-1}\\
{}&+\frac{(\delta_{1a_4}+\delta_{1a_5})(\delta_{2a_4}+\delta_{2a_5})}{\delta_{a_4a_5}-1}\bigg)+\frac{(p-2)(\delta_{12}-p)}{\delta_{12}-p+2}\bigg\}\;,
\end{split}
\end{equation}
\begin{equation}
\begin{split}
\mathcal{M}^{T, ({\rm II})}_{a_3a_4a_5}={}&-\frac{\sqrt{2}p(p-1)}{\delta_{a_4a_5}-1}\bigg\{\frac{(\delta_{2a_4}+\delta_{3a_4})(\delta_{2a_5}+\delta_{3a_5})}{\delta_{2a_3}-1}+\frac{(\delta_{1a_4}+\delta_{3a_4})(\delta_{1a_5}+\delta_{3a_5})}{\delta_{1a_3}-1}\\
{}&+\frac{2}{p(p-1)}\frac{(\delta_{1a_4}+\delta_{2a_4})(\delta_{1a_5}+\delta_{2a_5})}{\delta_{12}-p+1}+\frac{4(p-2)}{p(p-1)}\bigg(\frac{(\delta_{1a_4}+\delta_{2a_4}-1)(\delta_{1a_5}+\delta_{2a_5}-1)}{\delta_{12}-p+2}\\
{}&+\frac{p-3}{4}\frac{(\delta_{1a_4}+\delta_{2a_4}-2)(\delta_{1a_5}+\delta_{2a_5}-2)}{\delta_{12}-p+3}-\frac{1}{2}(p\delta_{a_4a_5}-p-1)\bigg)\bigg\}\;,
\end{split}
\end{equation}
\begin{equation}
\begin{split}
\mathcal{M}^{T, ({\rm III})}_{a_1a_2a_3a_4a_5}={}&-\frac{\sqrt{2}p(p-1)}{\delta_{a_2a_3}-1}\bigg\{\frac{(\delta_{a_1a_2}+\delta_{a_2a_5})(\delta_{a_1a_3}+\delta_{a_3a_5})}{\delta_{a_1a_5}-1}+\frac{(\delta_{a_1a_2}+\delta_{a_2a_4})(\delta_{a_1a_3}+\delta_{a_3a_4})}{\delta_{a_1a_4}-1}\\
{}&+\frac{2}{p(p-1)}\frac{(\delta_{a_2a_4}+\delta_{a_2a_5})(\delta_{a_3a_4}+\delta_{a_3a_5}+p-2)}{\delta_{a_4a_5}-1}-\frac{p-2}{p-1}\bigg(\frac{\delta_{a_1a_2}\delta_{a_2a_4}}{\delta_{a_1a_5}-1}+\frac{\delta_{a_1a_2}\delta_{a_2a_5}}{\delta_{a_1a_4}-1}\\
{}&-\frac{1+2p}{p}\delta_{a_1a_2}-\frac{\delta_{a_2a_3}}{p}+1\bigg)\bigg\}\;.
\end{split}
\end{equation}
They become $\mathcal{M}_1^T$ in (\ref{M1T}) when $p=2$. Finally, the coefficients of the two new structures $N^{({\rm I})}_{345}$, $N^{({\rm II})}_{a_3a_4a_5}$ are
\begin{equation}
\begin{split}
\mathcal{M}^{N, ({\rm I})}_{345}={}&\sqrt{2}p^2(p-2)\delta_{12}\bigg\{\frac{1}{(\delta_{15}-1)(\delta_{23}-1)}+\frac{1}{(\delta_{15}-1)(\delta_{24}-1)}+\frac{1}{(\delta_{13}-1)(\delta_{24}-1)}\\
{}&+\frac{1}{(\delta_{13}-1)(\delta_{25}-1)}+\frac{1}{(\delta_{14}-1)(\delta_{23}-1)}+\frac{1}{(\delta_{14}-1)(\delta_{25}-1)}\\
{}&-\frac{2}{p}\bigg(\frac{1}{\delta_{15}-1}+\frac{1}{\delta_{25}-1}+\frac{1}{\delta_{13}-1}+\frac{1}{\delta_{23}-1}+\frac{1}{\delta_{14}-1}+\frac{1}{\delta_{24}-1}\bigg)\bigg\}\;,
\end{split}
\end{equation}
\begin{equation}
\begin{split}
\mathcal{M}^{N, ({\rm II})}_{a_3a_4a_5}={}&-\sqrt{2}p(p-2)\delta_{12}\bigg\{\frac{\delta_{2a_3}}{(\delta_{1a_5}-1)(\delta_{2a_4}-1)}+\frac{\delta_{1a_4}}{(\delta_{2a_5}-1)(\delta_{1a_3}-1)}\\
{}&+\frac{1+\delta_{12}-p(\delta_{1a_3}+\delta_{2a_4}+\delta_{a_3a_4})}{p(\delta_{2a_4}-1)(\delta_{1a_3}-1)}\bigg\}\;.
\end{split}
\end{equation}
Note that they are proportional to $p-2$ and therefore vanish for $p=2$. 

Let us also make a comment regarding the seemingly confusing bevahior at the flat-space limit. The flat-space amplitude which one obtains from holographic correlators corresponds to that of gravitons. In general, one expects that the dependence on the KK levels should factorize as different KK modes all correspond to the same particle in flat space. However, this is not the case if we naively take the high energy limit of the Mellin amplitudes. Clearly, the $p$-dependence is not factored out as  the component amplitudes of the new R-symmetry structures for $p>2$ have the same high energy scaling behavior as the other component amplitudes. To understand this, it is important to note that the flat-space amplitude from AdS is in a special kinematic configuration where the polarizations of the gravitons are perpendicular to all the momenta \cite{Alday:2021odx}. However, such an amplitude for five points is zero in flat space.\footnote{This is easiest to see using double copy. The gluon five-point amplitude with orthogonal polarizations vanishes because it is impossible to contract five polarization vectors among themselves. By double copy, the graviton five-point amplitude also vanishes.} Therefore, the high energy limit of the Mellin amplitudes is not the flat-space amplitude as one might have naively expected. In fact, in applying the prescription of \cite{Penedones:2010ue}, there is an additional power of the inverse AdS radius $1/R$ which renders the flat-space limit zero. In other words, the high energy limit of the Mellin amplitudes computes only the $1/R$ corrections. We expect these corrections to have the same power counting for different KK modes. However, we do not expect their explicit expressions to be universal. 

\subsection{A comment on consistency}
Let us make a comment regarding the consistency of our result. In Section \ref{Sec:Mellin} we proved the truncation of the poles in $\delta_{12}$ by using factorization in the $(12)$ channel which only exploits the general analytic structure of the resulting four-point amplitude. Here we point out that the truncation can also be seen from a different point of view when it involves simultaneous poles with another channel. For concreteness, let us focus on the residue of the amplitude at the pole $\delta_{45}=1$. The residue is, via the factorization in the (45) channel, related to a four-point function $\langle pp 2 X\rangle$ where the first three operators are 1, 2, 3 respectively. As we know from (\ref{possibleexchanges}), the operator $X$ belongs to the $k=2$ multiplet and can be the superprimary $\mathcal{O}_2$, the R-symmetry current $\mathcal{J}_\mu$ or the stress tensor $\mathcal{T}_{\mu\nu}$. The Mellin amplitude of $\langle pp 2 X\rangle$ contains poles in $\delta_{12}$ due to the operator exchanges in the $(12)$ channel. These four-point Mellin amplitudes are given explicitly in Appendix \ref{Sec:strongcouplingcorr} and we observe a truncation of the subleading poles in $\delta_{12}$ for $m\geq 3$. This gives another derivation of the structure of the simultaneous poles in $\delta_{12}$ and $\delta_{45}$.  

Similar consistency checks have also been performed in other channels ({\it e.g.}, in the $(13)$ and $(45)$ channel), as well as for the R-symmetry  gluing (see Appendix \ref{Sec:rSymmetryGluing} for details).

\subsection{Comments on position space}\label{Sec:bootstrapposition}
Up to this point, all of our discussions are exclusively in Mellin space. This is mainly because of the simplified analytic structure of Mellin amplitudes, as can be seen from our main result (\ref{M5ptgeneralp}). However, it is also sometimes convenient to have position space expressions as some information is difficult to extract from the Mellin space representation. This has to do with the fact that certain nonzero expressions in position space may naively vanish in Mellin space. More precisely, different inverse Mellin transformations can only be added up if their contours can be smoothly deformed from one to another. Usually the contour part is ignored for simplicity and one just adds up the Mellin amplitudes. This causes some information to be lost in the process. In fact, we have already encountered such an example in this paper: The Drukker-Plefka twisted correlator is a constant in position space but has zero Mellin amplitude.\footnote{See also \cite{Rastelli:2017udc} for more examples in four-point functions.} The existence of the ambiguities makes a direct translation of Mellin space results into position space difficult.

One could also try to directly extend the position space algorithm of \cite{Goncalves:2019znr} to the $\langle pp222\rangle$ correlators. However, as explained in the introduction, this is technically difficult. Here we propose a hybrid approach. As explained in \cite{Goncalves:2019znr,Alday:2022lkk}, all five-point Witten diagrams can be expressed as a linear combinations of five-point $D$-functions by using integrated vertex identities\footnote{It is known for some time \cite{DHoker:1999mqo}  that four point  exchange Witten diagrams can be express in terms $D$-functions when certain conditions on the dimension of the operators are met, which is what often happens in $\mathcal{N}=4$ SYM.}. It is then  natural to construct an ansatz in  position space directly in terms of the $D$-functions. This will avoid directly computing the Witten diagrams which is a nontrivial task.  More concretely, we propose that the ansatz for $G_p$ in position space should have the following form
\begin{align}
A_{\Delta_1\dots \Delta_5}(x_i) = \sum_{\{\beta\}}c_{\{\beta\}}(t_{ij}) (x_{ij}^2)^{-\beta_{ij}} D_{\tilde{\Delta}_1\dots \tilde{\Delta}_5} (x_i)\;,\label{eq:positionspaceansatz}
\end{align}
where the coefficients $c_{\{\beta\}}(t_{ij})$ are linear combinations of all possible R-symmetry structures. The summation over $\beta_{ij}$ are subjected to the constraints 
\begin{align}
&\tilde{\Delta}_{i}+\sum_{j}\beta_{ij}  = \Delta_i \label{betaconstraints1},\\
&\sum_{i}\tilde{\Delta}_i\leq 2+\sum_{i} \Delta_i, \label{betaconstraints2}\\
& \beta_{ij}>0\;,\;\; \beta_{kl}>0\;,\quad \text{only if } \{i,j\}\neq \{k,l\}\label{betacompatible}\\
&\beta_{ij}\ge -2, \, \, \, \label{betaconstraints3}\\
&\beta_{12} \le p-1, \,\quad \beta_{ij} \leq 1\;\;(i,j\neq 1,2)\label{betaconstraints4}\;.
\end{align}
Let us now unpack these constraints a little. The first condition (\ref{betaconstraints1}) ensures that the external operators have the correct weights under conformal transformations. The constraint (\ref{betaconstraints2}) imposes a bound on the sum of weights in each $D$-functions.\footnote{One can see explicitly that it is the case for the $p=2$ five-point function. Moreover, the same bound also holds for four-point functions of higher KK modes.} This is expected if we use the integrated vertex identities\footnote{These will generalize the ones presented in Appendix A of \cite{Goncalves:2019znr} for $p=2$.} to reduce the exchange Witten diagrams to contact Witten diagrams. Exchanging single-trace operators leads to singularities in position space. The condition (\ref{betacompatible}) is the statement that particle exchanges have to be in the compatible channels. The constraint (\ref{betaconstraints3}) arises because the exchanged single-trace operator operators have maximal spin 2. To understand this more precisely, let us notice the following translation between position and Mellin space
\begin{align}
\prod_{1\leq i<j\leq 5}(x_{ij}^2)^{-\alpha_{ij}} D_{\tilde{\Delta}_1\dots \tilde{\Delta}_5} \rightarrow \mathcal{M}^{\alpha_{ij}}(\delta)= \frac{\pi^\frac{d}{2} \Gamma\left(\frac{\sum_i \tilde{\Delta}_i-d}{2}\right)}{\prod_{i}\Gamma(\tilde{\Delta}_i)} \prod_{i<j} \frac{\Gamma(\delta_{ij}-\alpha_{ij})}{\Gamma(\delta_{ij})}\;.\label{eq:fromPositionToMellin}
 \end{align}
The condition (\ref{betaconstraints3}) ensures in Mellin space that the numerator associated with an exchange pole is at most quadratic. Finally, the constraint (\ref{betaconstraints4}) controls the twists of the exchanged single-trace operators. Let us emphasize that this position ansatz, as it stands, does not manifest the truncation of poles seen in  (\ref{eq:ansatzMellin}). Nevertheless, this truncation can still be imposed in position space, though in a more intricate manner (this is in stark contrast with Mellin space). 
We notice that a given negative power  $(x_{12}^2)^{-\alpha}$  will lead to poles in Mellin space at all the locations $\delta_{12}=1,2,\dots ,\alpha$. Therefore, even though the $\delta_{12}$ poles in Mellin space truncate according to (\ref{eq:ansatzMellin}), in position space the result will necessarily involve all negative powers of $(x_{12}^2)^{-\alpha}$ with $\alpha=1,2,\ldots,p-1$. Truncation only implies that the negative powers are related but cannot just simply eliminate a subset of them. This is another instance where we can see explicitly that Mellin space is simpler.

To fix the coefficients in the ansatz, one can translate the ansatz back into Mellin space and compare with the Mellin amplitude (\ref{M5ptgeneralp}). This can be achieved by using the rule (\ref{eq:fromPositionToMellin}). However, as explained above, only some of the coefficients can be fixed due to the ambiguities of the translation. One may wonder if implementing the Drukker-Plefka twist and the chiral algebra condition in position space\footnote{See Appendix D of  \cite{Goncalves:2019znr} for more details on how to obtain explicit expressions for $D$-functions.} will give rise to additional constraints. But unfortunately we find that this is not the case. There still remains the possibility that one can fix the remaining coefficients using the recently derived higher-point lightcone conformal blocks \cite{Bercini:2021jti} to impose factorization in position space. But we have not found a  very efficient way to implement this. Therefore, we will postpone the task of finding the expressions in position space and leave it to future work.

%% file: sections/discussion.tex
\section{Discussions and outlook}\label{Sec:discussion}
In this paper we continued our journey of exploring the structure of five-point functions of $\frac{1}{2}$-BPS operators of 4d $\mathcal{N}=4$ SYM in the strongly coupled regime which is dual to $AdS_5\times S^5$ IIB supergravity. We improved the bootstrap approach of \cite{Goncalves:2019znr} which relies only on superconfromal symmetry and consistency with factorization. The important difference compared to the old approach is that both constraints are now implemented in Mellin space. Moreover, in the new method we only need to use the Drukker-Plefka twist and the chiral algebra condition is not needed. Using this approach, we obtained in a closed form the Mellin amplitudes for the infinite family of correlators of the form $\langle pp222 \rangle$. 

Compared to the simplest $\langle 22222\rangle$ case studied in \cite{Goncalves:2019znr}, the pole structure of the Mellin amplitudes of operators with higher KK levels is in general more complicated. However, an important simplifying feature we observed in this paper is a new type of pole truncation phenomenon. We find that the residues of certain poles associated with conformal descendants vanish. Morevoer, in the $\langle pp222\rangle$ case the number of poles does not grow with respect to $p$ when $p$ is large enough. Consequently, the pole structure of the Mellin amplitudes is much simpler than what is naively expected. This property played an important role in obtaining the $\langle pp222\rangle$ amplitudes and also gives us hope to bootstrap in closed forms more general families of five-point functions with different KK levels. 

Note that in deriving the pole truncation conditions, we have only used general properties of Mellin factorization. The same argument holds in many other theories and we expect similar simplifications in the pole structure. This leads to a number of possible extensions of our results in different setups. A prime example to consider is the gluon sector of certain 4d $\mathcal{N}=2$ SCFTs which is dual to SYM in $AdS_5\times S^3$. The first five-point function for the lowest KK level has been computed in \cite{Alday:2022lkk}. To make further progress in computing amplitudes of higher KK levels, one can adapt the strategy used here. One important ingredient which still needs to be worked out is the relations between different component correlators of the super four-point functions (see \cite{Bissi:2022wuh} for progress in this direction). This would be the input for exploiting the full power of the Mellin factorization. However, this will be a direct generalization to what we have done in Appendix \ref{Sec:higherkkSmult}. Another interesting application is to the 6d $\mathcal{N}=(2,0)$ theory which is dual to eleven dimensional supergravity in $AdS_7\times S^4$.

Going beyond five-point functions, an exciting future direction is to compute the super graviton six-point function of $AdS_5\times S^5$ IIB supergravity. This will provide a new benchmark for the program of holographic correlators at higher points. The results in this paper can already help us gain a nontrivial amount of knowledge of the structure of this new correlator. Moreover, much of the technology developed here, in particular the Mellin Drukker-Plefka twist, can also be straightforwardly applied to that problem. It appears to be a feasible target and we hope to report progress in this direction in the near future. 

Finally, let us mention that the $\langle pp222\rangle$ five-point functions we computed in this paper contain a wealth of new data of 4d $\mathcal{N}=4$ SYM. Through OPE, we can extract various non-protected three- and four-point functions. In \cite{Goncalves:2019znr} we constructed five-point conformal blocks (see \cite{Bercini:2020msp,Buric:2021ywo,Buric:2020dyz,Antunes:2021kmm,Fortin:2022grf} for progress in higher-point conformal blocks)  and explained how to use them to extract data from the $p=2$ five-point correlator. It would be interesting to perform a similar analysis here for the $\langle pp222\rangle$ correlators. The expression we have for general $p$ will be helpful for solving the mixing problem for the CFT data which is similar to the one appearing in four-point functions. It would also be interesting to extract the chiral algebra correlator from our supergravity result and compare with the field theory calculation. The four-point function case has been analyzed in \cite{Rastelli:2017ymc,Behan:2021pzk}.

%% file: sections/appendixKinematics.tex
\section{Higher R-charge super multiplet}\label{Sec:higherkkSmult}
A key element of the bootstrap analysis undertaken in the main text is the factorization of Mellin amplitudes into lower-point correlators. 
As explained in Section \ref{Subsec:Mellinfactorization} we do need as an input the explicit expression for the Mellin amplitudes associated with the four-point functions 
\begin{align}
\langle \mathcal{O}_2\mathcal{O}_2\mathcal{O}_2\mathcal{O}_2\rangle
\qquad
\langle \mathcal{J}_2\mathcal{O}_2\mathcal{O}_2\mathcal{O}_2\rangle
\qquad
\langle \mathcal{T}_2\mathcal{O}_2\mathcal{O}_2\mathcal{O}_2\rangle
\\
\langle \mathcal{O}_2\mathcal{O}_p\mathcal{O}_p\mathcal{O}_2\rangle
\qquad
\langle \mathcal{J}_2\mathcal{O}_p\mathcal{O}_p\mathcal{O}_2\rangle
\qquad
\langle \mathcal{T}_2\mathcal{O}_p\mathcal{O}_p\mathcal{O}_2\rangle
\\
\langle \mathcal{O}_p\mathcal{O}_p\mathcal{O}_2\mathcal{O}_2\rangle
\qquad
\langle \mathcal{J}_p\mathcal{O}_p\mathcal{O}_2\mathcal{O}_2\rangle
\qquad
\langle \mathcal{T}_p\mathcal{O}_p\mathcal{O}_2\mathcal{O}_2\rangle
\end{align}
where $\mathcal O_p$, $\mathcal{J}_p$ and  $\mathcal{T}_p$ denotes the following components of the half-BPS  supermultiplet $\mathbb{O}_p$ 
\begin{align}
\mathcal{O}_p\,:&\qquad \Delta=p\,,\,\,\,\,\,\,\,\,\,\,\,\,\,\mathcal{R}=[0,p,0]\,,\,\,\,\,\,\,\,\,\,\,\,\,\,\,\text{spin $0$}\,,
\\
\mathcal{J}_p\,:&\qquad \Delta=p+1\,,\,\,\,\mathcal{R}=[1,p-2,1]\,,\,\,\,\,\text{spin $1$}\,,
\\
\mathcal{T}_p\,: &\qquad  \Delta=p+2\,,\,\,\,\mathcal{R}=[0,p-2,0]\,,\,\,\,\,\text{spin $2$}\,.
\end{align}
In the special case $p=2$ they correspond respectively to the $\mathfrak{su}(4)$ current and stress tensor, hence their names.
The first goal of this appendix is to explain how the correlators above can be extracted from the $\langle \mathcal{O}_p\mathcal{O}_p\mathcal{O}_2\mathcal{O}_2\rangle$ component. 
This is a generalization of what has been done in \cite{Belitsky:2014zha} for the case $p=2$.
The second goal of this appendix is to explain how the factorization in Mellin space is implemented in the presence of some global symmetry. This is done in Appendix \ref{Sec:rSymmetryGluing}.
A final warning about notation is necessary. In the main text we use the six component null vectors on which the R-symmetry act linearly. Here, as it is natural from the super-space prospective will use four component R-symmetry variables $y$. The basic two-point invariants are identified as 
\begin{equation}
t_{ij}=y_{ij}^2\,.
\end{equation}
\subsection{Conventions}
In the following we will list all the 
conventions for raising and lowering indices 
\begin{equation}
y^{a \dot a} = \epsilon^{a b} y_{\dot b b} \epsilon^{\dot b \dot a}\,,
\end{equation}
where the $\epsilon$ tensor is defined with
\begin{equation}
\epsilon^{12}=\epsilon_{12}=1\,.
\end{equation}
It follows that
\begin{equation}
(y_{1i})_{ \dot a a} (y_{1j})^{a \dot a } = y_{1i}^2 + y_{1j}^2 - y_{ij}^2\,,
\end{equation}
which, in a particular case becomes
\begin{equation}
\det y_{ij} = \frac{1}{2} (y_{ij})_{ \dot a a} (y_{ij})^{a \dot a } = y_{ij}^2 \,.
\end{equation}
The Schouten identity can be used to show that
\begin{equation}
\epsilon^{\dot a \dot b} \epsilon^{ b a} y_{ij}^2 = y_{ij}^{a \dot a } y_{ij}^{b \dot b  }-y_{ij}^{b \dot a } y_{ij}^{a \dot b } \,.
\end{equation}
Finally, the inverse can easily be seen to be 
\begin{equation}
y_{ \dot a a }^{-1} = \frac{y_{\dot a a}}{y^2}\,,
\end{equation}
and, with these conventions we also have
\begin{equation}
\frac{\partial}{\partial y^{a \dot a }} y^2 = y_{ \dot a a}\,.
\end{equation}

\subsection{Differential Operators}
In order to consider different components of the $\frac{1}{2}$-BPS supermultiplets we will work in analytic superspace. The eight bosonic and eight fermionic coordinates of this superspace are packaged in a supermatrix
\begin{equation}
X^{\mathsf{A} \dot{\mathsf{A}}} = 
\begin{pmatrix}
x^{\alpha \dot \alpha } & \rho^{\alpha \dot  a} \\
\bar{\rho}^{a \dot \alpha} & y^{ a \dot a} 
\end{pmatrix}\,,
\end{equation}
whose superdeterminant is
\begin{equation}
\mathrm{sdet} X = \frac{\det \left(x^{\alpha \dot \alpha } - \rho^{ \alpha \dot a} y^{-1}_{\dot a  a} \bar{\rho} ^{ a \dot \alpha}\right) }{ \det y^{ a \dot a}}\,.
\end{equation}
The supersymmetrization of the propagator $d_{ij}= y_{ij}^2 / x_{ij}^2$ is given by
\begin{equation}\label{SuperD}
\hat d_{ij} = \frac{1}{\mathrm{sdet}(X_{ij})} = \frac{y_{ij}^2}{\hat x_{ij}^2}\,,
\end{equation}
where we introduce the short-hand notation
\begin{equation}
\hat x^{\alpha  \dot \alpha} = x^{ \alpha \dot  \alpha} - \rho^{ \alpha \dot a} y^{-1}_{\dot a  a} \bar{\rho} ^{ a \dot\alpha}\,.
\end{equation}
The two-point function of half-BPS superfields $\mathbb{O}_p$ is then simply
\begin{equation}
\langle \mathbb{O}_p (X_i) \mathbb{O}_p(X_j) \rangle = (\hat d_{ij})^p\,.
\end{equation}
The relevant superdescendants are obtained extracting the appropriate component by acting with  certain differential operators:
\begin{align}
\mathcal J_p&= \frac{1}{2} \mathcal D^{(J)}\mathbb{O}_p(X)\big{|}_{\rho=\bar{\rho}=0}
\,,\nonumber\\
\mathcal T_p &= \frac{1}{4} \mathcal D^{(T)}\mathbb{O}_p(X)\big{|}_{\rho=\bar{\rho}=0}\,.
\end{align}
Given the charges and symmetries of those operators the ansatz for the differential operators needs to be\footnote{These differential operators depend on $p$ through the coefficients $\mu$, $\nu_1$, $\nu_2$.
This dependence is not explicit in the notation.}
\begin{equation}\label{DJ}
\mathcal D^{(J)}=\lambda^{\alpha}\bar{\lambda}^{\dot{\alpha}}v^a\bar{v}^{\dot{a}}\Big{(} \frac{\partial}{\partial \bar{\rho}^{ a  \dot \alpha}} \frac{\partial}{\partial \rho^{ \alpha  \dot a}} + \mu \frac{\partial}{\partial y^{ a \dot a}} \frac{\partial}{\partial x^{ \alpha \dot \alpha}}\Big{)}\,,
\end{equation}
and
\begin{align}\label{DT}
\mathcal D^{(T)} =& \lambda^{\alpha_1}\lambda^{\alpha_2}\bar{\lambda}^{\dot{\alpha}_1}\bar{\lambda}^{\dot{\alpha}_2}
\epsilon^{\dot a_1 \dot a_2}\epsilon^{a_1a_2} \times \Big{(} 
\frac{\partial}{\partial \bar{\rho}^{a_1 \dot{\alpha}_1}} \frac{\partial}{\partial \bar{\rho}^{a_2 \dot \alpha_2}} \frac{\partial}{\partial \rho^{\alpha_1 \dot a_1}} \frac{\partial}{\partial \rho^{\alpha_2 \dot a_2}}+ \nonumber \\
&
+
\nu_1\, \frac{\partial}{\partial \bar{\rho}^{a_1 \dot \alpha_1}} 
\frac{\partial}{\partial\rho^{\alpha_1 \dot{a}_1}}
\frac{\partial}{\partial y^{a_2\dot a_2}} \frac{\partial}{\partial x^{\alpha_2 \dot \alpha_2}} 
+\nu_2\,  \frac{\partial}{\partial y^{a_1\dot a_1}} \frac{\partial}{\partial y^{a_2\dot a_2}} \frac{\partial}{\partial x^{\alpha_1 \dot \alpha_1}} \frac{\partial}{\partial x^{\alpha_2 \dot \alpha_2}}\Big{)}\,,
\end{align}
Before fixing the coefficients let us quote two simple identities which are very useful in the following\footnote{The second identity is obtained as follows
\begin{equation}
 0=   \frac{\partial}{\partial X^{\mathsf{A}\dot{\mathsf{A}}}}\,
\delta^{\mathsf{C}}_{\mathsf{B}}=
\frac{\partial}{\partial X^{\mathsf{A}\dot{\mathsf{A}}}}\,
X^{\mathsf{C} \dot{\mathsf{B}}}X^{-1}_{\dot{\mathsf{B}}\mathsf{B}}
=\delta^{\mathsf{C}}_{\mathsf{A}}\,\delta^{\dot{\mathsf{B}}}_{\dot{\mathsf{A}}}\,
X^{-1}_{\dot{\mathsf{B}}\mathsf{B}}
+(-1)^{(|\mathsf{A}|+|\dot{\mathsf{A}}|)(|\mathsf{C}|+|\dot{\mathsf{B}}|)}\,
X^{\mathsf{C} \dot{\mathsf{B}}}\,
\frac{\partial}{\partial X^{\mathsf{A}\dot{\mathsf{A}}}}\,
X^{-1}_{\dot{\mathsf{B}}\mathsf{B}}\,.
\end{equation}
Multiplying this equation by $(-1)^{(|\mathsf{A}|+|\dot{\mathsf{A}}|)(|\mathsf{C}|+|\dot{\mathsf{B}}|)}$ and $X^{-1}_{\dot{\mathsf{C}}\mathsf{C}}$ from the left (with summation over $\mathsf{C}$) we obtain \eqref{derivativeXinverse}.
}
\begin{equation}
\frac{\partial}{\partial X^{\mathsf{A}\dot{\mathsf{A}}}}
\frac{1}{\text{sdet}(X)}=-(-1)^{|\mathsf{A}|}\,\frac{X^{-1}_{\dot{\mathsf{A}}\mathsf{A}} }{\text{sdet}(X)}\,,
\end{equation}
\begin{equation}
\label{derivativeXinverse}
\frac{\partial}{\partial X^{\mathsf{A}\dot{\mathsf{A}}}}\,
X^{-1}_{\dot{\mathsf{B}}\mathsf{B}}\,=\,
-(-1)^{(|\mathsf{A}|+|\dot{\mathsf{A}}|)(|\mathsf{A}|+|\dot{\mathsf{B}}|)}\,
X^{-1}_{\dot{\mathsf{B}}\mathsf{A}}\,X^{-1}_{\dot{\mathsf{A}}\mathsf{B}}\,,
\end{equation}
where $|\alpha|=|\dot{\alpha}|=0\,,
\,\,\,\,
|a|=|\dot{a}|=1$.
In order to fix the coefficients in the ansatz \eqref{DJ}, \eqref{DT} it suffices to impose that two-point functions do not have off-diagonal components between different superdescendants. So we impose
\begin{equation}
\langle \mathcal J_p(1) \mathcal O_p (2) \rangle = \mathcal D^{(J)}_1 \langle \mathbb{O}_p (X_1) \mathbb{O}_p(X_2) \rangle \big|_{\rho,\bar{\rho}=0} \stackrel{!}{=} 0\,,
\end{equation}
which fixes the unknown coefficient in $\mathcal D^{(J)}$ to be
\begin{equation}\label{mu}
\mu=\frac{1}{p}\,.
\end{equation}
The action of the resulting operator on the two-point function is given by\footnote{The fact that is vanishes when $p=1$ is consistent with the fact that in this case the (field strength) supermultiplet is ultrashort and does not possess a $\mathcal{J}$ component.}
\begin{equation}
\mathcal{D}_1^{(J)}\, (\hat d_{12})^p=
(1-p^2)\,
\lambda_1^{\alpha}\,
\bar{\lambda}_1^{\dot{\alpha}}\,
v_1^a\,
\bar{v}_1^{\dot{a}}\,
X^{-1}_{\dot{\alpha} a}\,
X^{-1}_{\dot{a} \alpha}\,
 (\hat d_{12})^p\,,
\end{equation}
where $X=X_{12}$,
from which one derives the two-point function of the descendant $\mathcal{J}$ using the formula
\begin{equation}
\mathcal{D}_1^{(J)}\,\mathcal{D}_2^{(J)}\, (\hat d_{12})^p\big|_{\rho,\bar{\rho}=0}\,=\,
(1-p^2) 
\left(\bar{\lambda}_1 x_{12}^{-1}\lambda_2\right)
 \left(\bar{\lambda}_2 x_{12}^{-1}\lambda_1\right)
  \left(\bar{v}_1 y_{12}^{-1}v_2\right)
   \left(\bar{v}_2 y_{12}^{-1}v_1\right)
(d_{12})^p\,.
\end{equation}
From this equation we can extract the normalization of $\mathcal{J}_p$.
For the spin 2 operator we need to consider 
\begin{align}
\langle \mathcal T_p(1) \mathcal O_p (2) \rangle &= \mathcal D^{(T)}_1 \langle \mathbb{O}_p (X_1) \mathbb{O}_p(X_2) \rangle \big|_{\rho,\bar{\rho}=0}\stackrel{!}{=} 0 \,,\nonumber\\
\langle \mathcal T_p(1) \mathcal J_p (2) \rangle &= \mathcal D^{(T)}_1 \mathcal D^{(J)}_2 \langle \mathbb{O}_p (X_1) \mathbb{O}_p(X_2) \rangle \big|_{\rho,\bar{\rho}=0}\stackrel{!}{=}0  \,,
\end{align}
which in turn fixes the coefficients $\nu_i$ to be
\begin{equation}\label{nu}
\nu_1 = - \frac{4}{2+p} \,,\qquad 
\nu_2 = - \frac{2}{(1+p)(2+p)}\,.
\end{equation}
In the case of the stress tensor multiplet, when $p=2$, these coefficients agree with those found in \cite{Belitsky:2014zha}. 
The action of the resulting operator on the two-point function is given by
\begin{equation}
\mathcal{D}_1^{(T)}\, (\hat d_{12})^p=
2p^2(p-1)(p+3)
\lambda_1^{\alpha_1}\lambda_1^{\alpha_2}\bar{\lambda}_1^{\dot{\alpha}_1}\bar{\lambda}_1^{\dot{\alpha}_2}
\epsilon^{\dot a_1 \dot a_2}\epsilon^{a_1a_2} 
X_{\dot{\alpha}_1 a_1}^{-1}
X_{\dot{\alpha}_2 a_2}^{-1}
X_{\dot{a}_1 \alpha_1}^{-1}
X_{\dot{a}_2 \alpha_2}^{-1}
 (\hat d_{12})^p\,,
\end{equation}
where $X=X_{12}$,
from which one derives the two-point function of the descendant $\mathcal{T}$ using the formula
\begin{equation}
\mathcal{D}_1^{(T)}\,\mathcal{D}_2^{(T)}\, (\hat d_{12})^p\big|_{\rho,\bar{\rho}=0}=
16 p^2(p-1)(p+3)\,
\left(\bar{\lambda}_1 x_{12}^{-1}\lambda_2\right)^2
 \left(\bar{\lambda}_2 x_{12}^{-1}\lambda_1\right)^2
\frac{(y_{12}^2)^{p-2}}{(x_{12}^2)^{p+2}}\,.
\end{equation}
Three-point function with one descendant operator can be obtained using the formulae
\begin{align}
\mathcal{D}_1^{(J)}\, (\hat d_{12})^{a} (\hat d_{13})^{p-a}\big|_{\rho,\bar{\rho}=0}
&
=\,A
\,\Lambda_{1,23}
\,V_{1,23}\,
(d_{12})^{a} (d_{13})^{p-a}\,,
\\
\mathcal{D}_1^{(T)}\, (\hat d_{12})^{p} (\hat d_{13})^{p-a}\big|_{\rho,\bar{\rho}=0}
&
=\,B
\,
\left(\Lambda_{1,23}\right)^2
\,\det(y_{12}^{-1}-y_{13}^{-1})
(d_{12})^{a} (d_{13})^{p-a}\,,
\end{align}
where\begin{equation}
\label{eq:LambdaxandLambdaYdef}
\Lambda_{1,23}:=\bar{\lambda}_1 (x_{12}^{-1}-x_{13}^{-1})\lambda_1\,,
\qquad
V_{1,23}:=\bar{v}_1 (y_{12}^{-1}-y_{13}^{-1})v_1\,.
\end{equation}
and
\begin{equation}
A=-\frac{a(p-a)}{p}\,,
\qquad
B=-\frac{8a(a+1)(p-a)(p-a+1)}{(p+1)(p+2)}\,.
\end{equation}
\subsection{Four-point functions}
The two- and three-point functions of $\mathbb{O}_p$ operators are related in a simple way to the ones of their superprimaries  $\mathcal{O}_p$: they are obtained by replacing the propagators $d_{ij}$ with the super-propagators $\hat{d}_{ij}$. For four-point functions the situation is more involved due to the presence of cross ratios, but it is still true that the correlators of $\mathbb{O}_p$ is uniquely fixed by the one of $\mathcal{O}_p$. This is achieved by replacing the familiar space-time and R-symmetry cross ratios 
\begin{align}
u&= \frac{x_{12}^2 x_{34}^2}{x_{13}^2 x_{24}^2} = z \bar z\,, & v&= \frac{x_{14}^2 x_{23}^2}{x_{13}^2 x_{24}^2}=(1-z)(1-\bar z)\,, \nonumber\\
\sigma&= \frac{y_{12}^2 y_{34}^2}{y_{13}^2 y_{24}^2} = \alpha \bar \alpha\,, & \tau&= \frac{y_{14}^2 y_{23}^2}{y_{13}^2 y_{24}^2}=(1-\alpha)(1-\bar \alpha)\,.
\end{align}
with their super-symmetrizations, namely the four eigenvalues of the supermatrix
\begin{equation}
Z = X^{}_{12} X_{13}^{-1} X^{}_{34} X_{24}^{-1} \,.
\end{equation}
More explicitly,
we can extract the independent superconformal invariants by taking four independent supertraces
\begin{equation}
\label{superTraces}
\widehat{t}_k = \mathrm{Str} (Z^k)\,=\,
 \hat{z}^k + \hat{\bar z}^k - \hat{\alpha}^k -\hat{\bar\alpha}^k\,,
\qquad 
k=1,2,3,4\,.
\end{equation}
When all fermionic variables are set to zero the matrix above reduces to
\begin{equation}
Z \big|_{\rho,\bar{\rho}=0} = \begin{pmatrix}
x^{}_{12} x_{13}^{-1} x^{}_{34} x_{24}^{-1} &0 \\
0& y^{}_{12} y_{13}^{-1} y^{}_{34} y_{24}^{-1}
\end{pmatrix}\,.
\end{equation}
and upon taking the supertrace gives
\begin{equation}\label{t0}
t_k:=\widehat{t}_k \big|_{\rho,\bar{\rho}=0} =  z^k + \bar z^k - \alpha^k -\bar\alpha^k\,,
\end{equation}
which establishes the relation between the quantities $t_k$ and the cross ratios introduced above.
In terms of the cross-ratios the four point function reads
\begin{equation}
\label{superOfourpoints}
\langle \mathbb{O}_{p_1}(X_1) \mathbb{O}_{p_1}(X_2) \mathbb{O}_{p_2}(X_3) \mathbb{O}_{p_2}(X_4) \rangle = (\hat d_{12})^{p_1} \,(\hat d_{34})^{p_2}\; \mathcal G( \hat{z}, \hat{\bar z}; \hat{\alpha},\hat{\bar\alpha})\,.
\end{equation}
The function $\mathcal G$ satisfies the super-conformal Ward Identities and have a specific polynomial dependence on the R-symmetry cross ratios. We will come back to these constraints momentarily.
To extract the relevant components from \eqref{superOfourpoints} we need to act with the differential operators $\mathcal D^{(J)}$ and $\mathcal D^{(T)}$ given in \eqref{DJ}, \eqref{DT}.
\paragraph{Action of $\mathcal D^{(J)}$, $\mathcal D^{(T)}$ on four-point functions.}
 The spinning four-point functions are extracted by the action of the differential operators from \eqref{DJ} and \eqref{DT}
\begin{align}
\langle \mathcal J_{p_1}(1) \mathcal O_{p_1}(2) \mathcal O_{p_2}(3) \mathcal O_{p_2}(4) \rangle &= \tfrac{1}{2}\mathcal D_1^{(J)} \langle 
\mathbb{O}(X_1) \mathbb{O}_{p_1}(X_2) \mathbb{O}_{p_2}(X_3)\mathbb{O}_{p_2}(X_4) \rangle \big|_{\rho,\bar{\rho}=0} \,,\nonumber\\
\langle \mathcal T_{p_1}(1) \mathcal O_{p_1}(2) \mathcal O_{p_2}(3) \mathcal O_{p_2}(4) \rangle &= \tfrac{1}{4}\mathcal D_1^{(T)} \langle 
\mathbb{O}(X_1) \mathbb{O}_{p_1}(X_2) \mathbb{O}_{p_2}(X_3)\mathbb{O}_{p_2}(X_4)\rangle \big|_{\rho,\bar{\rho} =0} \,,
\label{Don4pt}
\end{align}
with coefficients determined in \eqref{mu} and \eqref{nu} above. 
In what follows we will always apply the differential operator at point $1$, so we will need to consider two particular cases of the four-point function, either $p_1=2$ and $p_2=p$, or the opposite.
The action of derivatives on the superpropagators are discussed in the previous section. 
The action of derivatives on the $\mathcal{G}$ factor is done in two steps. 
First we relate the derivatives with respect to the eigenvalues of the $Z$ matrix
\begin{equation}
z_1 = \hat{z}\,,\qquad z_2 = \hat{\bar z}\,,\qquad z_3 = \hat{\alpha}\,,\qquad z_4 = 
\hat{\bar \alpha}\,. 
\end{equation}
to derivatives with respect to the supertraces \eqref{superTraces}.
This is done by using the chain rule 
\begin{equation}
\frac{\partial \mathcal G}{\partial \widehat{t}_j} = 
\sum_{i=1}^4\,
\frac{\partial z_i}{\partial \widehat{t}_j}\,
\frac{\partial \mathcal G}{\partial z_i}\,.
\end{equation}
The Jacobian matrix can be derived easily since the variables are related according to \eqref{t0}, and is given by
\begin{equation}
\frac{\partial z_i}{\partial t_j} =\frac{(-1)^{j+F_i}}{j} \frac{Q^{(i)}_{4-j}}{\prod_{k\neq i}(z_i-z_k)}\,,
\end{equation}
where $Q^{(i)}_{4-j}$ are symmetric polynomials formed with the three variables $z_{k\neq i}$ (here written for $i=4$)
\begin{align}
Q^{(4)}_0 &=1 \,, & Q^{(4)}_1&= z_1 + z_2 + z_3 \,,\nonumber\\
Q^{(4)}_2 &=z_1 z_2 + z_1 z_3 + z_2 z_3  \,,& Q^{(4)}_3&= z_1  z_2  z_3\,,
\end{align}
and $F_1=F_2=0$ and $F_3=F_4=1$.
The second step is to take derivatives of $\widehat{t}_k$ with respect to the supercoordinates $ X_1^{\mathsf{A}\dot{\mathsf{A}}}$ using, for example
\begin{equation}
\frac{\partial}{\partial Z^{\mathsf{A}}_{\mathsf{B}}}
\widehat{t}_k =
k\,(-1)^{|\mathsf{A}|}\, (Z^{k-1})_{\mathsf{A}}^{\mathsf{B}}\,,
\end{equation}
\begin{equation}
\frac{\partial}{\partial X_1^{\mathsf{A}\dot{\mathsf{A}}}}
\widehat{t}_k =
k\,(-1)^{|\mathsf{A}|}\, 
(X_{12}^{-1}Z^{k}X_{12}^{})_{\dot{\mathsf{A}}}^{\dot{\mathsf{B}}}\,
(X_{12}^{-1}-X_{13}^{-1})_{\dot{\mathsf{B}} \mathsf{A}}\,,
\end{equation}
and similarly for higher derivatives.
This procedure is straightforward but tedious, the result takes the schematic form given in \eqref{JTpos}.

\paragraph{General structure of the correlator.}
Superconformal Ward identities and polynomiality in the R-symmetry variables imply that 
\begin{equation}\label{Factorized}
\langle \mathcal O_{p}(1) \mathcal O_{p}(2) \mathcal O_{2}(3) \mathcal O_{2}(4) \rangle =  G^{\mathrm{free}} + d_{12}^{p-2}\, R\, H_p(u,v)\,,
\end{equation}
where $R$ is the well-known function
\begin{align}
R ={}&v \,d_{12}^2 d_{34}^2  + \frac{v}{u} d_{13}^2 d_{24}^2  + \frac{v^2}{u} \,d_{14}^2 d_{23}^2  + \frac{v}{u}(v -u -1 ) d_{12} d_{13} d_{24} d_{34} \nonumber\\
&+\frac{v}{u}( 1- u-v) d_{12} d_{14} d_{23} d_{34}+ \frac{v}{u}(u-1-v)d_{13} d_{14} d_{23} d_{24} \,.
\end{align}
The free piece of the correlator can be supersymmetrized as shown in the next paragraph, while the supersymmetrization of the anomalous component is achieved with the method described above, where we supersymmetrize the cross ratios. The spinning anomalous functions will then be expressed in terms of derivatives of the dynamical function $H_p(u,v)$. 

\paragraph{The free theory check.}
As a check of the formulae derived in the previous section, will now consider the case of correlators in the free field theory. 
In the $SU(N)$ gauge theory, and for the particular configuration we are interested in, the tree-level four-point functions at any value of $N$ are
\begin{align}\label{Test1}
\langle \mathcal O_p(1) \mathcal O_p (2) \mathcal O_2 (3)\mathcal O_2(4) \rangle^{\text{free}} ={}&d_{12}^p d_{34}^2+\delta_{2p}\left(
d_{14}^2 d_{23}^2 + d_{13}^2 d_{24}^2\right)
+ \frac{2p(p-1)}{N^2-1} d_{12}^{p-2} d_{14} d_{23} d_{13} d_{24} \nonumber\\
&+ \frac{2p}{N^2-1} d_{12}^{p-1} d_{34} (d_{14} d_{23}+ d_{13} d_{24})
\,.
\end{align}
The four-point function $\langle \mathbb O_p \mathbb O_p \mathbb O_2\mathbb O_2\rangle$ is obtained from the above by simply replacing the propagator $d_{ij}$ with its supersymmetrized version $\hat d_{ij}$ introduces in \eqref{SuperD}.
We can rewrite this expression in terms of cross ratios as
\begin{equation}
\label{Gpp22}
G_{pp22} = 1 + \delta_{2p}\left(
\frac{v^2 \sigma^2}{u^2 \tau^2} + \frac{\sigma^2}{u^2}\right)+
\frac{2p}{N^2-1}\left((p-1)\frac{u^2\tau}{v\sigma^2} + \frac{u\tau }{v\sigma}+ \frac{u}{\sigma} \right)\,.
\end{equation}
In this case, the correlation function of superdescendants can be obtained either applying the general procedure discussed in the previous paragraph or by replacing the propagator $d_{ij}$ with $\hat d_{ij}$ in \eqref{Test1} and then applying the differential operators $\mathcal{D}^{(J)}$, $\mathcal{D}^{(T)}$. 
Both procedures give the same result, as they should, providing a check of the general procedure.

\paragraph{Frame simplifications.}
The computation we described can be simplified by choosing a frame. First, we wish only to apply the differential operator on the point $1$ of the four-point function, so we can set to zero the fermionic variables associated to the remaining points from the beginning. Second, the matrix $Z$ is superconformally invariant, so we can take advantage of conformal and $R$-symmetry transformations to send both $x_2$ and $y_2$ to 0, while sending $x_3$ and $y_3$ to infinity. Effectively the computation simplifies significantly to the evaluation of
\begin{equation}
\widehat{t}_k= \mathrm{Str} \left( (X_1^{} X_4^{-1})^k\right) \big|_{\rho_{i>1},\bar{\rho}_{i>1}=0}\,,
\end{equation}
where the matrix $Z$ becomes
\begin{equation}
\left(X^{}_1 X_4^{-1} \right)^{\mathsf{A}}_{\mathsf{B}} \big|_{\rho_4,\bar{\rho}_4=0}= 
\begin{pmatrix}
(x_1^{} x_4^{-1} )^{\alpha}_{\beta} & (\rho_1^{} y_4^{-1})^{ \alpha}_{b} \\
(\bar{\rho}^{}_1 x_4^{-1})^{a}_{\beta} & (y_1^{} y_4^{-1})^{ a}_{b} 
\end{pmatrix}\,,
\end{equation}
and the cross ratios in this frame are given by
\begin{align}
\frac{x_1^2}{x_4^2} &= z \bar z \,,&  \frac{x_{14}^2}{x_4^2}= (1-z)(1-\bar z) \,,\nonumber\\
\frac{y_1^2}{y_4^2} &= \alpha \bar \alpha \,,&  \frac{y_{14}^2}{y_4^2}= (1-\alpha)(1-\bar \alpha) \,.
\end{align}
With a  simple calculation we obtain (in this frame)
\begin{align}
\widehat{t}_1 &= t_1= \mathrm{Tr}(x^{}_1 x_4^{-1}) - \mathrm{Tr}(y^{}_1 y_4^{-1}) = z + \bar z - \alpha - \bar \alpha
\\
\widehat{t}_2 &= t_2 - 2 \;\mathrm{Tr}\left(\bar{\rho} x_4^{-1} \rho y_4^{-1}\right)\,, \nonumber \\
\widehat{t}_3 &= t_3 -3 \; \mathrm{Tr}\left(\bar{\rho} x_4^{-1} x_1^{} x_4^{-1} \rho y_4^{-1}\right)  -3 \; \mathrm{Tr}\left(\bar{\rho} x_4^{-1}\rho y_4^{-1} y_1^{} y_4^{-1} \right)\,,\nonumber \\
\widehat{t}_4 &= t_4  -4 \; \mathrm{Tr}\left(\bar{\rho} x_4^{-1} (x_1^{} x_4^{-1})^2 \rho y_4^{-1}\right)  -4 \; \mathrm{Tr}\left(\bar{\rho} x_4^{-1}\rho y_4^{-1} (y_1^{} y_4^{-1})^2 \right) \nonumber\\
& \hspace{17.5mm} -4 \; \mathrm{Tr}\left(\bar{\rho} x_4^{-1} x_1^{} x_4^{-1} \rho y_4^{-1}  y_1^{} y_4^{-1}\right) -2 \; \mathrm{Tr}\left(\bar{\rho} x_4^{-1} \rho y_4^{-1}\bar{\rho} x_4^{-1} \rho y_4^{-1}\right)\,,
\end{align}
where $\rho=\rho_1$, $\bar{\rho}=\bar{\rho}_1$.

\paragraph{Summary.}
The final expression for the spinning correlators in \eqref{Don4pt} involves the structures  $\Lambda_{1,ij}$ and $V_{1,ij}$ introduced in \eqref{eq:LambdaxandLambdaYdef}. These quantities are not independent but satisfy the relation 
\begin{equation}\label{relation}
\Lambda_{1,24}= \Lambda_{1,23} + \Lambda_{1,34}\,,
\end{equation}
and similarly for $V_{1,ij}$.
In particular the correlator involving $\mathcal{J}_p$ is linear in $\Lambda_{1,ij}$ and $H_{1,ij}$,
while the one involving $\mathcal{T}_p$ is quadratic in $\Lambda_{1,ij}$ and independent of $H_{1,ij}$.
Once the general expression for the correlator is obtained in terms of $\Lambda_{1,ij}$, one can decompose into 
\begin{equation}
\bar{\lambda}_1x_{1k}^{-1}\lambda_1=\frac{z \cdot x_{1k}}{x_{1k}^2}\,,
\qquad
z^{\mu}=\sigma^{\mu}_{\alpha \dot{\alpha}}\lambda_1^{\alpha}\bar{\lambda}_1^{\dot \alpha}
\end{equation}
 elements, which will have a natural counterpart in the Mellin approach of the next section
 (compare to \eqref{scalardeltaLR})
\begin{align}\label{JTpos}
\langle \mathcal J_{p_1}(1) \mathcal O_{p_1}(2) \mathcal O_{p_2}(3) \mathcal O_{p_2}(4)\rangle &= \frac{1}{x_{12}^{2 p_1} x_{34}^{2p_2}} \sum_{k=2}^4   \alpha^{(k)}_{p_1,p_2}(u,v; y_{ij},Y_{1,ij})\frac{z \cdot x_{1k}}{x_{1k}^2}\,,\nonumber\\
\langle \mathcal T_{p_1}(1) \mathcal O_{p_1}(2) \mathcal O_{p_2}(3) \mathcal O_{p_2}(4)\rangle &= \frac{1}{x_{12}^{2p_1} x_{34}^{2p_2}} \sum_{k,l=2}^4   \beta_{p_1,p_2}^{(k,l)}(u,v; y_{ij})\frac{z \cdot x_{1k}}{x_{1k}^2}\frac{z \cdot x_{1l}}{x_{1l}^2}\,,
\end{align}
where 
\begin{equation}
Y_{1,ij}=  y_{1i}^2 y_{1j}^2\, V_{1,ij}\,.
\end{equation}

\subsection{R- Symmetry gluing}\label{Sec:rSymmetryGluing}

\paragraph{Realization of $\mathfrak{su}(4)$ R-symmetry in the space of polynomials.}
It is convenient to use an index free notation to implement finite dimensional representations of $\mathfrak{su}(4)$. The components of a given representation are packaged in a polynomial $\mathsf{O}_{\mathcal{R}}(y,v,\bar{v})$ in the variables $y^{a\dot a}, v^a, \bar{v}^{\dot{a}}$ (here $a \in \{1,2\}, \dot{a} \in \{\dot{1},\dot{2}\}$) subject to certain constraints that depend on the $\mathfrak{su}(4)$ Dynkin labels $\mathcal{R}=[q,p,r]$.
The fist constraint states that $\mathsf{O}_{\mathcal{R}}(y,v,\bar{v})$ is homogeneous in $v$ and $\bar{v}$ of degree $q$ and $r$ respectively. The second constraint is slightly more involved. In the case $\mathcal{R}=[0,p,0]$, so that $\mathsf{O}_{\mathcal{R}}$ is independent of $v,\bar{v}$ it reads
\begin{equation}
\left(
w^a \bar{w}^{\dot{a}}\frac{\partial}{\partial y^{a\dot{a}}}\right)^{p+1}\mathsf{O}_{\mathcal{R}}(y)=0\,,
\qquad
\forall\,\,\,\,w,\bar{w}\,.
\end{equation}
The case $\mathcal{R}=[1,p-2,1]$ is more involved. Since we will not use it in this work we will not present the identification of 
$\mathcal{R}=[1,p-2,1]$ as the kernel of differential operators.  Two-point functions take the form
\begin{equation}
\label{2ptRsymmG}
G_{[q,p,r]}(1,2)=
(y^2_{12})^p\, (v_1 y_{12} \bar{v}_2)^q\,(v_2 y_{12} \bar{v}_1)^r\,.
\end{equation}

\paragraph{Projections and \emph{gluing}.}
To implement factorization in Mellin space in the presence of some global symmetry (in our case the $\mathfrak{su}(4)$ R-symmetry)
it is necessary to take into account this extra structure. To do so, we introduce a projector that singles out the contribution of a given operator\footnote{Here we use the notation $\mathsf{O}$ instead of $\mathcal{O}$ since we are ignoring the space-time part.} $\mathsf{O}$ which we denote by
\begin{equation}
\label{ProjectionORsymm}
|\mathsf{O}|= \frac{1}{\mathcal{N}_{\mathsf{O}}}\,\,
\mathcal{D}^{(\ell,r)}_{\mathcal{R}[{\mathsf{O}]}}
\,|\mathsf{O}(\ell)\rangle \langle \mathsf{O}^*(r)|\,\,\Big{|}_{\ell=r}\,,
\end{equation}
where $\mathcal{D}$ is a differential operator which is fixed (up to a normalization that will be explained momentarily) by the requirement that \eqref{ProjectionORsymm} is invariant under $\mathfrak{su}(4)$. The notation ${}^*$ denotes conjugation which acts on representations as $[q,p,r]^*=[r,p,q]$. When we insert the quantity  $|\mathsf{O}|$  in an n-point correlation function it is understood that we first place $|\mathsf{O}(\ell)\rangle \langle \mathsf{O}^*(r)|$, next act with the differential operator $\mathcal{D}$ on the coordinates $\ell$ and $r$ and finally set the coordinates $\ell$ and $r$ to be equal. To fix the normalization of $\mathcal{D}$ we insert $|\mathsf{O}|$ in the two-point function
\begin{equation}
\langle  \mathsf{O}^*(1) \mathsf{O}(2)\rangle \,=\,\mathcal{N}_{\mathsf{O}}\,G_{\mathcal{R}[\mathsf{O}]}(1,2)\,,
\end{equation}
where $G_{\mathcal{R}}$ is given in \eqref{2ptRsymmG} and obtain the condition
\begin{equation}
\label{eq:DGGisG}
\mathcal{D}^{(\ell,r)}_{\mathcal{R}}
G_{\mathcal{R}}(1,\ell)
G_{\mathcal{R}}(r,2)
\,\Big{|}_{\ell=r}\,=\,
G_{\mathcal{R}}(1,2)\,.
\end{equation}
The explicit form of $\mathcal{D}_{\mathcal{R}}$ is slightly complicated. The simplest one is given by
\begin{equation}
\mathcal{D}_{[0,p,0]}^{(\ell,r)}
=\sum_{n=0}^p \frac{\left(-\partial_{\ell}\cdot \partial_r\right)^{p-n}}{\Gamma(p+1)\Gamma(p+2)}
\sum_{k=0}^n\,
(-1)^{n-M(k,n-k)}
\frac{(p-n+1)_{m(k,n-k)+1}}{\Gamma(m(k,n-k)+1)}
(\square_\ell)^k
(\square_r)^{n-k}\,,
\end{equation}
where $(a)_n$ denotes the Pochhammer symbol,  $M(a,b)=\text{max}(a,b)$, $m(a,b)=\text{min}(a,b)$
and
\begin{equation}
\partial_{i}\cdot \partial_j:=
\epsilon^{a_1 a_2}\epsilon^{\dot{a}_1\dot{a}_2}
\,
\frac{\partial}{\partial y_{i}^{a_1\dot{a}_1}}
\frac{\partial}{\partial y_{j}^{a_2\dot{a}_2}}\,,
\qquad 
\square_i=\frac{1}{2}\,\partial_{i}\cdot \partial_{i}\,.
\end{equation}
The general expression for the differential operator $\mathcal{D}_{[1,p-2,1]}$ is more complicated, but it easy to obtain for fixed $p$ using the defining relation \eqref{eq:DGGisG}.
Let us report the simplest member of this family as an example
\begin{equation}
\mathcal{D}_{[1,0,1]}^{(\ell,r)}
=
\left(\partial_{v_{\ell}}\partial_{v_{r}}\right)
\left(\partial_{\bar{v}_{\ell}}\partial_{\bar{v}_{r}}\right)
\left(\tfrac{1}{2}\partial_{\ell}\cdot \partial_r
-\square_\ell-\square_r\right)
-\frac{3}{16}
\left(\partial_{v_{\ell}}\partial_{y_{\ell}}\partial_{\bar{v}_{\ell}}\right)
\left(\partial_{v_{r}}\partial_{y_{r}}\partial_{\bar{v}_{r}}\right)\,,
\end{equation}
where the contraction of indices is understood using the $\epsilon$ tensor.

\paragraph{Application to five-point functions.}
When we insert the projector \eqref{ProjectionORsymm} in a 5-point function we will produce a product of a 3-point and a 4-point function on which the differential operator $\mathcal{D}$ acts. In the following we denote by $\rightarrow$ the combination of acting with $\mathcal{D}^{(\ell,r)}$ and setting the coordinates $\ell=r$. 
The case that is relevant for the exchange of $\mathcal{O}_p$ which transform in a $[0,p,0]$ representation is 
\begin{equation}
\big{[}
(y_{1\ell}^2)\,(y_{2\ell}^2)^{p-1}
\big{]}
\,
\big{[}(y_{ri}^2)^{p-2}(y_{rj}^2)^{}(y_{rk}^2)^{}
\big{]}
\rightarrow
\label{eq:gluingScalar}
\end{equation}

\begin{equation}
\tfrac{1}{p}
(y_{2i}^2)^{p-3}
\left(
y^2_{2i}\,(y^2_{2j}y^2_{1k}+y^2_{1j}y^2_{2k })
+(p-2)y^2_{1i} y^2_{2j}y^2_{2 k}
-\tfrac{p-2}{p+1}\,
y^2_{12}\,(y^2_{2 k}y^2_{ij}+y^2_{2j}y^2_{ik})
-\tfrac{1}{p+1}\,
y^2_{12}
y^2_{2i}
y^2_{jk}
\right)
\end{equation}
Similarly, using the definitions above, gluing the 3 and 5 point functions corresponding to the exchange of $\mathcal{J}_p$ (which transforms in the representation $[1,p-2,1]$) is achieved by the subsitution
\begin{equation}
\Big{[}
(y_{2\ell}^2)^{p-2}\,Y_{\ell,12}\,\Big{]}
\Big{[}\,(y_{ri}^2)^{p-3}(y_{rj}^2)^{}\,Y_{r,kl}
\Big{]}
\rightarrow \end{equation}
\begin{equation}
(y_{2i}^2)^{p-4}\left(
y^2_{2i}\,y^2_{2j}\,(y^2_{1k}y^2_{2l}-y^2_{1l}y^2_{2k })
+y^2_{12}\left(
\tfrac{p-3}{p+2}
y^2_{2j}
(y^2_{il}y^2_{2k}-y^2_{2l}y^2_{ik })
+\tfrac{1}{p+2}
y^2_{2i}
(y^2_{jl}y^2_{2k}-y^2_{2l}y^2_{jk })
\right)
\right)
\end{equation}
For the exchange of $\mathcal{T}_p$ we use the same rules as \eqref{eq:gluingScalar} with $p$ replaced by $p-2$.

%% file: sections/appendixMellin.tex
\section{Strong coupling correlators}
\label{Sec:strongcouplingcorr}

We can define the inverse Mellin transform of the scalar correlator as 
\begin{equation}
\langle \mathcal O_{p_1}(1) \mathcal O_{p_1}(2) \mathcal O_{p_2}(3) \mathcal O_{p_2}(4) \rangle =  d_{12}^{p_1} \, d_{34}^{p_2}\; G(z_k)= \int \mathrm d \delta_{ij} \mathcal{M}(\delta_{ij},y_{ij})\prod_{i<j} \frac{\Gamma(\delta_{ij})}{x_{ij}^{2\delta_{ij}}}\,.
\end{equation}
Conformal symmetry requires the Mellin variables $\delta_{ij}$ to obey the following equations 
\begin{equation}
\sum_{j\neq i} \delta_{ij} = \Delta_i\,,
\end{equation}
effectively leaving only two degrees of freedom for four-point functions. It is useful to consider the following parametrization
\begin{equation}\label{sij1}
\delta_{ij}= \frac{\Delta_i + \Delta_j - s_{ij}}{2}\,,
\end{equation}
so that the solution is given simply as
\begin{equation}\label{sij2}
s_{12}=s_{34}=s \,,\qquad  s_{14}= s_{23} = t\,,\qquad s_{13}=s_{24}= 2 (p_1+p_2) - s -t\,.
\end{equation}
For the configuration we are interested in we can then write the inverse Mellin transform as
\begin{equation}
G(u,v;\sigma,\tau) = \int \frac{\mathrm d s \mathrm d t}{4} u^{\frac{s}{2}} v^{\frac{t-p_1-p_2}{2}}  \mathcal{M}(s,t;\sigma,\tau) \prod_{i<j} \Gamma(\delta_{ij}(s,t))\,.
\end{equation}
Equivalently, the Mellin transform of the spacetime correlator is
\begin{equation}
\mathcal{M}(s,t;\sigma,\tau) \prod_{i<j} \Gamma(\delta_{ij}(s,t)) = \int_0^\infty \mathrm{d} u \int_0^\infty \mathrm{d}v \;u^{-\frac{s}{2}-1} v^{\frac{p_1+p_2-t}{2}-1} G(u,v;\sigma,\tau) \,.
\end{equation}
When the correlator has a factorized form as in \eqref{Factorized}, then it is convenient to introduce the Mellin transform of the dynamical function $\mathcal H_p(u,v)$
\begin{equation}
\widetilde{\mathcal{M}}_p(s,t) \prod_{i<j} \Gamma(\tilde\delta_{ij}(s,t)) = \int_0^\infty \mathrm{d} u \int_0^\infty \mathrm{d}v \;u^{-\frac{s}{2}-1} v^{\frac{p-t}{2}} H_p(u,v) \,,
\end{equation}
where the shifted variables are defined as
\begin{align}
\tilde\delta_{13} &= \delta_{13}+2 \,,\qquad \tilde \delta_{24}=\delta_{24}+2\,,\nonumber\\
\delta_{ij}&=\delta_{ij} \quad\mathrm{otherwise,}
\end{align}
and make crossing properties of the Mellin amplitude simpler.
At strong coupling the Mellin space version of the correlator was found to have a particularly simple structure \cite{Rastelli:2016nze,Rastelli:2017udc}, and in the case under consideration it reduces to
\begin{equation}
\widetilde{\mathcal{M}}_p(s,t) = \frac{32}{(s-2)(t-p)(p-s-t)}\,.
\end{equation}
For the spinning correlators we can also write inverse Mellin transforms as follows 
\begin{align}
\langle \mathcal J_{p_1}(1) \mathcal O_{p_1}(2) \mathcal O_{p_2}(3) \mathcal O_{p_2}(4) \rangle &= \sum_{k=2}^4 \frac{z \cdot x_{1k}}{x_{1k}^2}\int \left[\mathrm d \delta\right] \mathcal{M}_{p_1,p_2}^{k} \prod_{i=2}^4 \frac{\Gamma(\delta_i + \textrm{{\fancy{$\delta$}}}^k_i)}{x_{1i}^{2\delta_i}}\prod_{i<j} \frac{\Gamma(\delta_{ij})}{x_{ij}^{2\delta_{ij}}}\,,\nonumber\\
\langle \mathcal T_{p_1}(1) \mathcal O_{p_1}(2) \mathcal O_{p_2}(3) \mathcal O_{p_2}(4) \rangle &= \sum_{k,l=2}^4 \frac{z \cdot x_{1k}}{x_{1k}^2}\frac{z \cdot x_{1l}}{x_{1l}^2}\int \left[\mathrm d \delta\right] \mathcal{M}_{p_1,p_2}^{kl} \prod_{i=2}^4 \frac{\Gamma(\delta_i + \textrm{{\fancy{$\delta$}}}^k_i+\textrm{{\fancy{$\delta$}}}^l_i)}{x_{1i}^{2\delta_i}}\prod_{i<j} \frac{\Gamma(\delta_{ij})}{x_{ij}^{2\delta_{ij}}}\,,
\end{align}
with $\textrm{{\fancy{$\delta$}}}^k_i$ the Kronecker-delta, and the Mellin variables are constrained by
\begin{equation}
\delta_i= - \sum_{j=2}^4 \delta_{ij}\,,\qquad \delta_{ii}= - \Delta_i\,, \qquad \sum_{i,j=2}^4 \delta_{ij}= S-\Delta_1 \,.
\end{equation} 
In the two cases of interest we have $S-\Delta_1=p_1$, so the $\delta_{ij}$ variables have the same solution as in the scalar case, see \eqref{sij1} and \eqref{sij2}.
Comparing with the form of the correlators obtained in the previous section, we can see that the inverse Mellin trasform of the functions introduced in \eqref{JTpos} are exactly the $\mathcal{M}^{k}$ and $\mathcal{M}^{kl}$ above
\begin{align}
&\alpha_{p_1,p_2}^{(k)}(u,v;y_{ij},Y_{1,ij}) = \int \frac{\mathrm d s \mathrm d t}{4} u^{\frac{s}{2}} v^{\frac{t-p_1-p_2}{2}}  \mathcal{M}_{p_1,p_2}^{k}(s,t;y_{ij},Y_{1,ij})  \prod_{i=2}^4 \Gamma(\delta_i + \textrm{{\fancy{$\delta$}}}^k_i) \prod_{i<j}  \Gamma(\delta_{ij}) \,, \nonumber\\
&\beta_{p_1,p_2}^{(k,l)}(u,v;y_{ij}) = \int \frac{\mathrm d s \mathrm d t}{4} u^{\frac{s}{2}} v^{\frac{t-p_1-p_2}{2}}  \mathcal{M}_{p_1,p_2}^{kl}(s,t;y_{ij})  \prod_{i=2}^4 \Gamma(\delta_i + \textrm{{\fancy{$\delta$}}}^k_i + \textrm{{\fancy{$\delta$}}}^l_i) \prod_{i<j}  \Gamma(\delta_{ij}) \,.
\end{align}
Inversing the logic we then have 
\begin{align}
\mathcal{M}_{p_1,p_2}^{k}(s,t;y_{ij}, Y_{1,ij}) \prod_{i=2}^4 \Gamma(\delta_i + \textrm{{\fancy{$\delta$}}}^k_i) \prod_{i<j}  \Gamma(\delta_{ij}) &= \int_0^\infty \mathrm{d} u \, \mathrm{d}v \;u^{-\frac{s}{2}-1} v^{\frac{p_1+p_2-t}{2}-1} \alpha_{p_1,p_2}^{(k)}(u,v;y_{ij}, Y_{1,ij} ) \,, \nonumber\\
\mathcal{M}_{p_1,p_2}^{kl}(s,t;y_{ij}) \prod_{i=2}^4 \Gamma(\delta_i + \textrm{{\fancy{$\delta$}}}^k_i +\textrm{{\fancy{$\delta$}}}^l_i) \prod_{i<j}  \Gamma(\delta_{ij}) &= \int_0^\infty \mathrm{d} u \, \mathrm{d}v \;u^{-\frac{s}{2}-1} v^{\frac{p_1+p_2-t}{2}-1} \beta_{p_1,p_2}^{(k,l)}(u,v;y_{ij}) \,.
\end{align}
As explained in the previous section, the functions $\alpha^{(k)}_{p_1,p_2}$ and $\beta^{(k,l)}_{p_1,p_2}$ are given in terms of derivatives of the dynamical function from the scalar correlator. When $p_1=2$ and $p_2=p$,  or $p_1=p$ and $p_2=2$, we are then relating with $H_p$ from \eqref{Factorized}, and so we should use
\begin{align}
&\int_0^\infty \mathrm d u \int_0^\infty \mathrm d v \;u^{-\frac{s}{2}-1} v^{\frac{p+2-t}{2}-1} u^m v^n \frac{\partial^a}{\partial u^a} \frac{\partial^b}{\partial v^b} H_p(u,v) =\widetilde{\mathcal{M}}_p(s-2m+2a,t-2n+2b)  \nonumber\\
&\times  (-1)^{a+b} \left(m-a-\frac{s}{2}\right)_a \left(n-b+\frac{p+2-t}{2}\right)_b\prod_{i<j} \Gamma(\tilde \delta_{ij}(s-2m-2a,t-2n-2b))\,,
\end{align}
which allows us to write $\mathcal{M}_{p_1,p_2}^{k}$ and $\mathcal{M}_{p_1,p_2}^{kl}$ for those two configurations in terms of the scalar Mellin amplitude $\widetilde{\mathcal{M}}_p(s,t)$.
At the end of the day, the Mellin amplitudes for $\langle \mathcal J_2 \mathcal O_2 \mathcal O_p \mathcal O_p \rangle$ are
\begin{align}
\mathcal{M}_{2,p}^{2} &=  -2(t-p-2)\left(\frac{2(p-2)}{s-4} + \frac{2}{s-2}+ \frac{p}{4+p-s-t}\right)y_{24}^2\, y_{34}^{2(p-1)}\, Y_{1,23}  \nonumber\\
&\quad- 2(2+p-s-t)\left(\frac{2(p-2)}{s-4}+ \frac{2}{s-2}+\frac{p}{t-p}\right)y_{23}^2 \,y_{34}^{2(p-1)} \,Y_{1,24}  \nonumber\\
&\quad - 2p(s-2p) \left(\frac{1}{t-p} - \frac{1}{4+p-s-t}\right) y_{23}^2 \,y_{24}^2\, y_{34}^{2(p-2)} \,Y_{1,34} \,, \nonumber
\end{align}
\begin{align}
\mathcal{M}_{2,p}^{3} &= 2(t-p-2)\left(\frac{2}{s-2}+ \frac{p}{4+p-s-t}\right) y_{24}^2\, y_{34}^{2(p-1)}\, Y_{1,23}  \nonumber\\
&\quad+ 2(2+p-s-t)\left(\frac{2}{s-2}-\frac{p}{t-p}\right)y_{23}^2 \,y_{34}^{2(p-1)}\, Y_{1,24}  \nonumber\\
&\quad- 2p(s-2p) \left(\frac{1}{t-p}+\frac{1}{4+p-s-t}\right)y_{23}^2 \,y_{24}^2\, y_{34}^{2(p-2)} \,Y_{1,34} \label{eq:currentMellinexample} \,.
\end{align}
Note that in general we expected poles at $s-2$, $t-p$ and $p+4-s-t$. However, in the $\mathcal{M}_{2,p}^{2}$ component we see also the presence of a pole at $s-4$. While this might appear unexpected at first, it is in fact due to the shift in the Gamma functions of spinning correlators. When $p_1=2$ and $p_2=p$ the relevant factors are
\begin{equation}
\Gamma(\delta_2 + 1) \Gamma(\delta_{34}) = \Gamma\left(3-\frac{s}{2}\right) \Gamma\left(p-\frac{s}{2}\right)\,.
\end{equation}
It is then evident that the Gamma functions do not prohibit the satellite pole at $s-4$ (unless $p=2$, in which case the residue vanishes).
Meanwhile for $\langle \mathcal J_p \mathcal O_p \mathcal O_2 \mathcal O_2 \rangle$ we have
\begin{align}
\mathcal{M}_{p,2}^{2} &=  \frac{2(p-2)s}{p} \left[ \frac{t-p-2}{4+p-s-t} y_{12}^{2(p-3)} y_{13}^2 \,y_{24}^4 \,  Y_{1,23} +  \frac{2+p-s-t}{t-p} y_{12}^{2(p-3)} y_{14}^2 \,y_{23}^4 \,  Y_{1,24}\right. \nonumber\\
&\quad\left.+  2 \left(1 + \frac{p}{t-p} + \frac{p}{4+p-s-t}\right) y_{12}^{2(p-3)} y_{14}^2 \,y_{23}^2 \, y_{24}^2 \, Y_{1,23}\right] \nonumber\\
&\quad+ \frac{2(t-p-2)}{p}\left(p-2-\frac{4}{s-2}- \frac{2p}{4+p-s-t}\right) y_{12}^{2(p-2)} y_{24}^2 \,y_{34}^2  \, Y_{1,23} \nonumber\\
&\quad+ \frac{2(2+p-s-t)}{p}\left(p-2-\frac{4}{s-2}- \frac{2p}{t-p}\right)y_{12}^{2(p-2)} y_{23}^2 \,y_{34}^2  \, Y_{1,24} \nonumber\\
&\quad+ \frac{2}{p} \left(s (p-2) - \frac{2p(s-2p)}{t-p} + \frac{2p(s (p-1) - 2p)}{4+p-s-t}\right)\;y_{12}^{2(p-2)} y_{23}^2 \,y_{24}^2 \, Y_{1,34}\,, \nonumber
\end{align}
\begin{align}
\mathcal{M}_{p,2}^{3} &= \frac{2(p-2)(s-2p)}{p} \left[ \frac{t-p-2}{4+p-s-t} y_{12}^{2(p-3)} y_{13}^2 \,y_{24}^4 \,  Y_{1,23} +  \frac{2+p-s-t}{t-p} y_{12}^{2(p-3)} y_{14}^2 \,y_{23}^4 \,  Y_{1,24} \right.\nonumber\\
&\quad \left.+  2\left(1 + \frac{p}{t-p} + \frac{p}{4+p-s-t}\right) y_{12}^{2(p-3)} y_{14}^2 \,y_{23}^2 \, y_{24}^2 \, Y_{1,23} \right]\nonumber\\
&\quad+ \frac{2(t-p-2)}{p}\left(p-2+\frac{4(p-1)}{s-2}+ \frac{2p(p-1)}{4+p-s-t}\right) y_{12}^{2(p-2)} y_{24}^2 \,y_{34}^2  \, Y_{1,23} \nonumber\\
&\quad+ \frac{2(2+p-s-t)}{p}\left(p-2+\frac{4(p-1)}{s-2} - \frac{2p}{t-p}\right)y_{12}^{2(p-2)} y_{23}^2 \,y_{34}^2  \, Y_{1,24} \nonumber\\
&\quad+\frac{2(s-2p)}{p}\left(p-2-\frac{2p}{t-p} - \frac{2p}{4+p-s-t}\right) y_{12}^{2(p-2)} y_{23}^2 \,y_{24}^2 \, Y_{1,34}\,.
\end{align}
In this case the Gamma factors for $\mathcal{M}_{p,2}^{2}$ are
\begin{equation}
\Gamma(\delta_2 + 1) \Gamma(\delta_{34}) = \Gamma\left(p+1-\frac{s}{2}\right) \Gamma\left(2-\frac{s}{2}\right)\,,
\end{equation}
and that is why the shift does not lead to any unexpected pole. For the other Mellin components $\mathcal{M}_{p_1,p_2}^{3}$ and $\mathcal{M}_{p_1,p_2}^{4}$ we have
\begin{align}
\Gamma(\delta_3 + 1) \Gamma(\delta_{24}) &= \Gamma\left(\frac{s+t-p}{2}\right) \Gamma\left(\frac{s+t-p-2}{2}\right)\,,\nonumber\\
\Gamma(\delta_4 +1) \Gamma(\delta_{23}) &=  \Gamma\left(\frac{4+p-t}{2}\right) \Gamma\left(\frac{2+p-t}{2}\right)\,,
\end{align}
which explains why there cannot be any new poles in these channels for any of the two configurations considered.

Moving on to the spin 2 case, the Mellin amplitudes for the  $\langle \mathcal T_2 \mathcal O_2 \mathcal O_p \mathcal O_p \rangle$ correlator are
\begin{align}
\mathcal{M}_{2,p}^{2,2} &=  \frac{16}{3}\left(1-p+6(p-2)\left(\frac{p-3}{s-6} +\frac{2}{s-4}\right) + \frac{2}{s-2}+ \frac{p(p-1)}{t-p} + \frac{p(p-1)}{4+p-s-t}\right) y_{23}^2 \,y_{24}^2 \, y_{34}^{2(p-1)} \,, \nonumber
\end{align}
\begin{align}
\mathcal{M}_{2,p}^{2,3} &=  \frac{16}{3}\left(1-p-\frac{6(p-2)}{s-4}-\frac{4}{s-2}+ \frac{p(p-1)}{t-p}-\frac{2p(p-1)}{4+p-s-t}\right)y_{23}^2 \,y_{24}^2 \, y_{34}^{2(p-1)} \,, \nonumber\\
\mathcal{M}_{2,p}^{3,3} &=  \frac{16}{3}\left(1-p+\frac{2}{s-2}+ \frac{p(p-1)}{t-p}+\frac{p(p-1)}{4+p-s-t}\right)y_{23}^2 \,y_{24}^2 \, y_{34}^{2(p-1)} \label{eq:StressMellinexample}\,.
\end{align}
There are once again some satellite poles, but the explanation follows exactly the same reasoning as before. The relevant Gamma factors in $\mathcal{M}_{2,p}^{2,2}$ are in this case
\begin{equation}
\Gamma(\delta_2 + 2) \Gamma(\delta_{34}) = \Gamma\left(4-\frac{s}{2}\right) \Gamma\left(p-\frac{s}{2}\right)\,,
\end{equation}
thus allowing poles both at $s-4$ and $s-6$ (except if $p=2,3$). Meanwhile, for $\mathcal{M}^{2,3}$ (and also $\mathcal{M}^{2,4}$) the relevant Gammas are
\begin{equation}
\Gamma(\delta_2 + \textrm{{\fancy{$\delta$}}}^2_2) \Gamma(\delta_{34}) = \Gamma\left(3-\frac{s}{2}\right) \Gamma\left(p-\frac{s}{2}\right)\,,
\end{equation}
and so the only satellite pole in those Mellin components is at $s-4$.
At last, for the correlator $\langle \mathcal T_p \mathcal O_p \mathcal O_2 \mathcal O_2 \rangle$ we have
\begin{align}
\mathcal{M}_{p,2}^{2,2} &=  \frac{8(p-2)(s+2)}{(p+1)(p+2)}\left[\left(\frac{s(p-1)-2p}{t-p}+\frac{2p}{4+p-s-t}\right) y_{12}^{2(p-3)} \, y_{14}^2\, y_{23}^4 \,y_{24}^2 \right. \nonumber\\
&\qquad\qquad\qquad\qquad  \left.+\left(\frac{2p}{t-p}+\frac{s(p-1)-2p}{4+p-s-t}\right) y_{12}^{2(p-3)} \, y_{13}^2\, y_{23}^2 \,y_{24}^4 \right] \nonumber\\
&\quad +\frac{8 \,y_{12}^{2(p-2)} y_{23}^2\, y_{24}^2 \,y_{34}^2}{(p+1)(p+2)} \left((p-1)(p-2)s - 4(p^2-2)+\frac{16}{s-2}\right.\nonumber\\
&\qquad\qquad\qquad\qquad\qquad  \left.-\frac{2p(s(p-2)-2p)}{t-p}-\frac{2p(s(p-2)-2p)}{4+p-s-t}  \right)\,, \nonumber
\end{align}
\begin{align}
\mathcal{M}_{p,2}^{2,3} &= \frac{8 (p-2)}{(p+1)(p+2)}\left[\left(\frac{s^2(p-1)-2p^2 s-2(p+2)(p-1)}{t-p}-\frac{p(s(p-1)+6p+2)}{4+p-s-t}\right) y_{12}^{2(p-3)} \, y_{14}^2\, y_{23}^4 \,y_{24}^2 \right.  \nonumber\\
&\qquad\qquad\qquad\qquad\left.  +\left(\frac{2p(s-p+1)}{t-p}+\frac{s^2(p-1)-s p (p-1)+2(p^2+p+2)}{4+p-s-t}\right) y_{12}^{2(p-3)} \, y_{13}^2\, y_{23}^2 \,y_{24}^4 \right] \nonumber\\
&\quad +\frac{8\, y_{12}^{2(p-2)} \, y_{23}^2\, y_{24}^2 \,y_{34}^2 }{(p+1)(p+2)}\left((p-1)(p-2)s-2(p+2)(p-1)-\frac{16p}{s-2}\right.\nonumber\\
&\qquad\qquad\qquad  \left.-\frac{2p(s(p-2)-(p-1)(p+2))}{t-p}+\frac{p(s(p-1)(p-2)-2(p^2+p+2))}{4+p-s-t}\right) \,, \nonumber
\end{align}
\begin{align}
\mathcal{M}_{p,2}^{3,3} &= \frac{8 (p-2)(s-2p)}{(p+1)(p+2)}\left[\left(\frac{2p^2-(s-2)(p-1)}{t-p}-\frac{2p^2}{4+p-s-t}\right) y_{12}^{2(p-3)} \, y_{14}^2\, y_{23}^4 \,y_{24}^2 \right. \nonumber\\
&\qquad\qquad\qquad\qquad  \left.+\left(\frac{2p}{t-p}+\frac{s(p-1)+2}{4+p-s-t}\right) y_{12}^{2(p-3)} \, y_{13}^2\, y_{23}^2 \,y_{24}^4 \right] \nonumber\\
&\quad+ \frac{8\, y_{12}^{2(p-2)} \, y_{23}^2\, y_{24}^2 \,y_{34}^2 }{(p+1)(p+2)} \left((p-1)(p-2)s-4p+\frac{8p(p-1)}{s-2}\right.\nonumber\\
&\qquad\qquad\qquad\qquad\qquad  \left. +\frac{2p(2(p^2-2)-s(p-2))}{t-p}+\frac{2p^2(s(p-2)+2)}{4+p-s-t}\right) \,. \nonumber\\
\end{align}
Note that in the final expressions above we omit the $\mathcal{M}_{p_1,p_2}^{4}$ and $\mathcal{M}_{p_1,p_2}^{k,4}$ cases, but they can be easily obtained from the equations relating different Mellin components
\begin{align}
\sum_k \delta_{k} \mathcal{M}^{k} &= 0\,, \nonumber\\
\sum_k (\delta_k + \textrm{{\fancy{$\delta$}}}^l_k) \mathcal{M}^{kl} &=0 \,,
\end{align}
which play a similar role to the equation \eqref{relation} relating the tensor structures in position space. Finally, note that for the particular case of $p_1=p_2=2$ the expressions above simplify and agree with those found in our earlier work \cite{Goncalves:2019znr}.

\subsection{Example of factorization}
The goal of this subsection is to show explicitly how to use factorization, lower-point Mellin amplitudes and the $R$-symmetry gluing rules from Appendix \ref{Sec:rSymmetryGluing} to recover part of the five-point function. To simplify the presentation we will focus on the factorization of the scalar $\textrm{{\bf{20}}}'$ operator exchanged in the channel $(45)$. 

The building blocks for the factorization are the Mellin amplitude of the four-point function $\langle\mathcal{O}_p\mathcal{O}_p \mathcal{O}_2\mathcal{O}_2\rangle$ and the three-point function $\langle \mathcal{O}_2\mathcal{O}_2 \mathcal{O}_2 \rangle$
\begin{align}
\mathcal{M}_{pp22} =& \frac{4 t_{01} t_{23} t_{12}^{p-2} \left(\delta _{12} \left(p \left(t_{02} t_{13}-t_{03} t_{12}\right)-(p-1) t_{01} t_{23}\right)+(p-1) p t_{03} t_{12}+\delta _{12}^2 t_{01} t_{23}\right)}{\delta _{23}-1}+\dots\nonumber\\
\mathcal{M}_{222}=& \,C_{OOO}\,t_{45}t_{40}t_{50}
\end{align}
where we decided to write explicitly only part of the four point function to simplify even further the analysis. The label $0$ in the formula is associated to the operator that is being exchanged in the factorization channel. 

Now we can borrow the formula from (\ref{eq:factorizationequation},\ref{eq:QmScalarDefinition})to obtain
\begin{align}
\mathcal{M}_{pp222} = 2\Gamma(2) \frac{ \mathcal{M}_{pp22} \mathcal{M}_{222}}{(2\delta_{45}-2)} + \dots
\end{align}
where the $\dots$ stand for other poles and contributions of other operators. 
The gluing in R-symmetry space gives, implementing\footnote{Recall that $t_{ij}=y_{ij}^2$.}  (\ref{eq:gluingScalar}) for $p=2$,
\begin{align}
t_{\ell 4}t_{\ell 5}
\,t_{ri_1}t_{ri_2} &\rightarrow \frac{1}{2}\left((t_{4i_1}t_{5i_2}+t_{4i_2}t_{5i_1})-\frac{t_{45}t_{i_1i_2}}{3}\right)\;,\\
t_{\ell 4}t_{\ell 5}
t_{ri_1}^2 &\rightarrow  \, t_{4i_1}t_{5i_1}\;.
\end{align}
Thus we obtain
\begin{align}
&\mathcal{M}_{pp222} = \frac{2 C_{OOO} t_{23} t_{45} t_{12}^{p-2} }{3 \left(\delta _{23}-1\right) \left(\delta _{45}-1\right)} \big[\left(3 \delta _{12} \left(p\, t_{15} \left(t_{13} t_{24}-t_{12} t_{34}\right)+t_{14} \left(p \left(t_{13} t_{25}-t_{12} t_{35}\right)\right.\right.\right.\nonumber\\
&\left.\left.\left.-2 (p-1) t_{15} t_{23}\right)\right)+(p-1) p t_{12} \left(3 t_{15} t_{34}+3 t_{14} t_{35}-t_{13} t_{45}\right)+6 \delta _{12}^2 t_{14} t_{15} t_{23}\right) \big]+\dots
\end{align}
where, again, the dots stand for other poles and contributions of other operators. In particular this formula can be compared with our previous result for five point function of {\bf{20'}} operators.

%% file: higherweight5pt.bbl
\providecommand{\href}[2]{#2}\begingroup\raggedright\begin{thebibliography}{10}

\bibitem{Goncalves:2019znr}
V.~Gon{\c c}alves, R.~Pereira, and X.~Zhou, ``{$20'$ Five-Point Function from
  $AdS_5\times S^5$ Supergravity},''
  \href{http://dx.doi.org/10.1007/JHEP10(2019)247}{{\em JHEP} {\bfseries 10}
  (2019) 247},
\href{http://arxiv.org/abs/1906.05305}{{\ttfamily arXiv:1906.05305 [hep-th]}}.

\bibitem{Rastelli:2016nze}
L.~Rastelli and X.~Zhou, ``{Mellin amplitudes for $AdS_5\times S^5$},''
  \href{http://dx.doi.org/10.1103/PhysRevLett.118.091602}{{\em Phys. Rev.
  Lett.} {\bfseries 118} no.~9, (2017) 091602},
  \href{http://arxiv.org/abs/1608.06624}{{\ttfamily arXiv:1608.06624
  [hep-th]}}.

\bibitem{Rastelli:2017udc}
L.~Rastelli and X.~Zhou, ``{How to Succeed at Holographic Correlators Without
  Really Trying},'' \href{http://dx.doi.org/10.1007/JHEP04(2018)014}{{\em JHEP}
  {\bfseries 04} (2018) 014},
\href{http://arxiv.org/abs/1710.05923}{{\ttfamily arXiv:1710.05923 [hep-th]}}.

\bibitem{Bissi:2022mrs}
A.~Bissi, A.~Sinha, and X.~Zhou, ``{Selected topics in analytic conformal
  bootstrap: A guided journey},''
  \href{http://dx.doi.org/10.1016/j.physrep.2022.09.004}{{\em Phys. Rept.}
  {\bfseries 991} (2022) 1--89},
  \href{http://arxiv.org/abs/2202.08475}{{\ttfamily arXiv:2202.08475
  [hep-th]}}.

\bibitem{Alday:2020lbp}
L.~F. Alday and X.~Zhou, ``{All Tree-Level Correlators for M-theory on $AdS_7
  \times S^4$},'' \href{http://dx.doi.org/10.1103/PhysRevLett.125.131604}{{\em
  Phys. Rev. Lett.} {\bfseries 125} no.~13, (2020) 131604},
  \href{http://arxiv.org/abs/2006.06653}{{\ttfamily arXiv:2006.06653
  [hep-th]}}.

\bibitem{Alday:2020dtb}
L.~F. Alday and X.~Zhou, ``{All Holographic Four-Point Functions in All
  Maximally Supersymmetric CFTs},''
  \href{http://dx.doi.org/10.1103/PhysRevX.11.011056}{{\em Phys. Rev. X}
  {\bfseries 11} no.~1, (2021) 011056},
  \href{http://arxiv.org/abs/2006.12505}{{\ttfamily arXiv:2006.12505
  [hep-th]}}.

\bibitem{Rastelli:2019gtj}
L.~Rastelli, K.~Roumpedakis, and X.~Zhou, ``{$\mathbf{AdS_3\times S^3}$
  Tree-Level Correlators: Hidden Six-Dimensional Conformal Symmetry},''
  \href{http://dx.doi.org/10.1007/JHEP10(2019)140}{{\em JHEP} {\bfseries 10}
  (2019) 140},
\href{http://arxiv.org/abs/1905.11983}{{\ttfamily arXiv:1905.11983 [hep-th]}}.

\bibitem{Giusto:2020neo}
S.~Giusto, R.~Russo, A.~Tyukov, and C.~Wen, ``{The CFT$_6$ origin of all
  tree-level 4-point correlators in AdS$_3 \times S^3$},''
  \href{http://dx.doi.org/10.1140/epjc/s10052-020-8300-4}{{\em Eur. Phys. J. C}
  {\bfseries 80} no.~8, (2020) 736},
  \href{http://arxiv.org/abs/2005.08560}{{\ttfamily arXiv:2005.08560
  [hep-th]}}.

\bibitem{Alday:2021odx}
L.~F. Alday, C.~Behan, P.~Ferrero, and X.~Zhou, ``{Gluon Scattering in AdS from
  CFT},'' \href{http://dx.doi.org/10.1007/JHEP06(2021)020}{{\em JHEP}
  {\bfseries 06} (2021) 020}, \href{http://arxiv.org/abs/2103.15830}{{\ttfamily
  arXiv:2103.15830 [hep-th]}}.

\bibitem{Alday:2022lkk}
L.~F. Alday, V.~Gon\c{c}alves, and X.~Zhou, ``{Supersymmetric Five-Point Gluon
  Amplitudes in AdS Space},''
  \href{http://dx.doi.org/10.1103/PhysRevLett.128.161601}{{\em Phys. Rev.
  Lett.} {\bfseries 128} no.~16, (2022) 161601},
  \href{http://arxiv.org/abs/2201.04422}{{\ttfamily arXiv:2201.04422
  [hep-th]}}.

\bibitem{Aharony:2016dwx}
O.~Aharony, L.~F. Alday, A.~Bissi, and E.~Perlmutter, ``{Loops in AdS from
  Conformal Field Theory},''
  \href{http://dx.doi.org/10.1007/JHEP07(2017)036}{{\em JHEP} {\bfseries 07}
  (2017) 036},
\href{http://arxiv.org/abs/1612.03891}{{\ttfamily arXiv:1612.03891 [hep-th]}}.

\bibitem{Farrow:2018yni}
J.~A. Farrow, A.~E. Lipstein, and P.~McFadden, ``{Double copy structure of CFT
  correlators},'' \href{http://dx.doi.org/10.1007/JHEP02(2019)130}{{\em JHEP}
  {\bfseries 02} (2019) 130}, \href{http://arxiv.org/abs/1812.11129}{{\ttfamily
  arXiv:1812.11129 [hep-th]}}.

\bibitem{Armstrong:2020woi}
C.~Armstrong, A.~E. Lipstein, and J.~Mei, ``{Color/kinematics duality in
  AdS$_{4}$},'' \href{http://dx.doi.org/10.1007/JHEP02(2021)194}{{\em JHEP}
  {\bfseries 02} (2021) 194}, \href{http://arxiv.org/abs/2012.02059}{{\ttfamily
  arXiv:2012.02059 [hep-th]}}.

\bibitem{Albayrak:2020fyp}
S.~Albayrak, S.~Kharel, and D.~Meltzer, ``{On duality of color and kinematics
  in (A)dS momentum space},''
  \href{http://dx.doi.org/10.1007/JHEP03(2021)249}{{\em JHEP} {\bfseries 03}
  (2021) 249}, \href{http://arxiv.org/abs/2012.10460}{{\ttfamily
  arXiv:2012.10460 [hep-th]}}.

\bibitem{Jain:2021qcl}
S.~Jain, R.~R. John, A.~Mehta, A.~A. Nizami, and A.~Suresh, ``{Double copy
  structure of parity-violating CFT correlators},''
  \href{http://dx.doi.org/10.1007/JHEP07(2021)033}{{\em JHEP} {\bfseries 07}
  (2021) 033}, \href{http://arxiv.org/abs/2104.12803}{{\ttfamily
  arXiv:2104.12803 [hep-th]}}.

\bibitem{Zhou:2021gnu}
X.~Zhou, ``{Double Copy Relation in AdS Space},''
  \href{http://dx.doi.org/10.1103/PhysRevLett.127.141601}{{\em Phys. Rev.
  Lett.} {\bfseries 127} no.~14, (2021) 141601},
  \href{http://arxiv.org/abs/2106.07651}{{\ttfamily arXiv:2106.07651
  [hep-th]}}.

\bibitem{Diwakar:2021juk}
P.~Diwakar, A.~Herderschee, R.~Roiban, and F.~Teng, ``{BCJ amplitude relations
  for Anti-de Sitter boundary correlators in embedding space},''
  \href{http://dx.doi.org/10.1007/JHEP10(2021)141}{{\em JHEP} {\bfseries 10}
  (2021) 141}, \href{http://arxiv.org/abs/2106.10822}{{\ttfamily
  arXiv:2106.10822 [hep-th]}}.

\bibitem{Cheung:2022pdk}
C.~Cheung, J.~Parra-Martinez, and A.~Sivaramakrishnan, ``{On-shell correlators
  and color-kinematics duality in curved symmetric spacetimes},''
  \href{http://dx.doi.org/10.1007/JHEP05(2022)027}{{\em JHEP} {\bfseries 05}
  (2022) 027}, \href{http://arxiv.org/abs/2201.05147}{{\ttfamily
  arXiv:2201.05147 [hep-th]}}.

\bibitem{Herderschee:2022ntr}
A.~Herderschee, R.~Roiban, and F.~Teng, ``{On the differential representation
  and color-kinematics duality of AdS boundary correlators},''
  \href{http://dx.doi.org/10.1007/JHEP05(2022)026}{{\em JHEP} {\bfseries 05}
  (2022) 026}, \href{http://arxiv.org/abs/2201.05067}{{\ttfamily
  arXiv:2201.05067 [hep-th]}}.

\bibitem{Drummond:2022dxd}
J.~M. Drummond, R.~Glew, and M.~Santagata, ``{BCJ relations in ${AdS}_5 \times
  S^3$ and the double-trace spectrum of super gluons},''
  \href{http://arxiv.org/abs/2202.09837}{{\ttfamily arXiv:2202.09837
  [hep-th]}}.

\bibitem{Bissi:2022wuh}
A.~Bissi, G.~Fardelli, A.~Manenti, and X.~Zhou, ``{Spinning correlators in
  $\mathcal{N} = 2$ SCFTs: Superspace and AdS amplitudes},''
  \href{http://arxiv.org/abs/2209.01204}{{\ttfamily arXiv:2209.01204
  [hep-th]}}.

\bibitem{Armstrong:2022jsa}
C.~Armstrong, H.~Gomez, R.~Lipinski~Jusinskas, A.~Lipstein, and J.~Mei, ``{New
  recursions for tree-level correlators in (Anti) de Sitter space},''
  \href{http://arxiv.org/abs/2209.02709}{{\ttfamily arXiv:2209.02709
  [hep-th]}}.

\bibitem{Lee:2022fgr}
H.~Lee and X.~Wang, ``{Cosmological Double-Copy Relations},''
  \href{http://arxiv.org/abs/2212.11282}{{\ttfamily arXiv:2212.11282
  [hep-th]}}.

\bibitem{Li:2022tby}
Y.-Z. Li, ``{Flat-space structure of gluon and graviton in AdS},''
  \href{http://arxiv.org/abs/2212.13195}{{\ttfamily arXiv:2212.13195
  [hep-th]}}.

\bibitem{Caron-Huot:2018kta}
S.~Caron-Huot and A.-K. Trinh, ``{All tree-level correlators in
  AdS$_{5}$$\times$S$_{5}$ supergravity: hidden ten-dimensional conformal
  symmetry},'' \href{http://dx.doi.org/10.1007/JHEP01(2019)196}{{\em JHEP}
  {\bfseries 01} (2019) 196}, \href{http://arxiv.org/abs/1809.09173}{{\ttfamily
  arXiv:1809.09173 [hep-th]}}.

\bibitem{Goncalves:2014rfa}
V.~Gon{\c c}alves, J.~Penedones, and E.~Trevisani, ``{Factorization of Mellin
  amplitudes},'' \href{http://dx.doi.org/10.1007/JHEP10(2015)040}{{\em JHEP}
  {\bfseries 10} (2015) 040},
\href{http://arxiv.org/abs/1410.4185}{{\ttfamily arXiv:1410.4185 [hep-th]}}.

\bibitem{Beem:2013sza}
C.~Beem, M.~Lemos, P.~Liendo, W.~Peelaers, L.~Rastelli, and B.~C. van Rees,
  ``{Infinite Chiral Symmetry in Four Dimensions},''
  \href{http://dx.doi.org/10.1007/s00220-014-2272-x}{{\em Commun. Math. Phys.}
  {\bfseries 336} no.~3, (2015) 1359--1433},
  \href{http://arxiv.org/abs/1312.5344}{{\ttfamily arXiv:1312.5344 [hep-th]}}.

\bibitem{Drukker:2009sf}
N.~Drukker and J.~Plefka, ``{Superprotected n-point correlation functions of
  local operators in N=4 super Yang-Mills},''
  \href{http://dx.doi.org/10.1088/1126-6708/2009/04/052}{{\em JHEP} {\bfseries
  04} (2009) 052}, \href{http://arxiv.org/abs/0901.3653}{{\ttfamily
  arXiv:0901.3653 [hep-th]}}.

\bibitem{Bercini:2020msp}
C.~Bercini, V.~Gon\c{c}alves, and P.~Vieira, ``{Light-Cone Bootstrap of Higher
  Point Functions and Wilson Loop Duality},''
  \href{http://dx.doi.org/10.1103/PhysRevLett.126.121603}{{\em Phys. Rev.
  Lett.} {\bfseries 126} no.~12, (2021) 121603},
  \href{http://arxiv.org/abs/2008.10407}{{\ttfamily arXiv:2008.10407
  [hep-th]}}.

\bibitem{Mack:2009mi}
G.~Mack, ``{D-independent representation of Conformal Field Theories in D
  dimensions via transformation to auxiliary Dual Resonance Models. Scalar
  amplitudes},'' \href{http://arxiv.org/abs/0907.2407}{{\ttfamily
  arXiv:0907.2407 [hep-th]}}.

\bibitem{Penedones:2010ue}
J.~Penedones, ``{Writing CFT correlation functions as AdS scattering
  amplitudes},'' \href{http://dx.doi.org/10.1007/JHEP03(2011)025}{{\em JHEP}
  {\bfseries 03} (2011) 025}, \href{http://arxiv.org/abs/1011.1485}{{\ttfamily
  arXiv:1011.1485 [hep-th]}}.

\bibitem{Fitzpatrick:2011ia}
A.~L. Fitzpatrick, J.~Kaplan, J.~Penedones, S.~Raju, and B.~C. van Rees, ``{A
  Natural Language for AdS/CFT Correlators},''
  \href{http://dx.doi.org/10.1007/JHEP11(2011)095}{{\em JHEP} {\bfseries 11}
  (2011) 095}, \href{http://arxiv.org/abs/1107.1499}{{\ttfamily arXiv:1107.1499
  [hep-th]}}.

\bibitem{Zhou:2017zaw}
X.~Zhou, ``{On Superconformal Four-Point Mellin Amplitudes in Dimension
  $d>2$},'' \href{http://dx.doi.org/10.1007/JHEP08(2018)187}{{\em JHEP}
  {\bfseries 08} (2018) 187},
\href{http://arxiv.org/abs/1712.02800}{{\ttfamily arXiv:1712.02800 [hep-th]}}.

\bibitem{Zhou:2018ofp}
X.~Zhou, ``{On Mellin Amplitudes in SCFTs with Eight Supercharges},''
  \href{http://dx.doi.org/10.1007/JHEP07(2018)147}{{\em JHEP} {\bfseries 07}
  (2018) 147},
\href{http://arxiv.org/abs/1804.02397}{{\ttfamily arXiv:1804.02397 [hep-th]}}.

\bibitem{Goncalves:2014ffa}
V.~Gon\c{c}alves, ``{Four point function of $\mathcal{N}=4$ stress-tensor
  multiplet at strong coupling},''
  \href{http://dx.doi.org/10.1007/JHEP04(2015)150}{{\em JHEP} {\bfseries 04}
  (2015) 150}, \href{http://arxiv.org/abs/1411.1675}{{\ttfamily arXiv:1411.1675
  [hep-th]}}.

\bibitem{DHoker:1999mqo}
E.~D'Hoker, D.~Z. Freedman, and L.~Rastelli, ``{AdS / CFT four point functions:
  How to succeed at z integrals without really trying},''
  \href{http://dx.doi.org/10.1016/S0550-3213(99)00526-X}{{\em Nucl. Phys.}
  {\bfseries B562} (1999) 395--411},
\href{http://arxiv.org/abs/hep-th/9905049}{{\ttfamily arXiv:hep-th/9905049
  [hep-th]}}.

\bibitem{Bercini:2021jti}
C.~Bercini, V.~Gon\c{c}alves, A.~Homrich, and P.~Vieira, ``{The Wilson loop
  \textemdash{} large spin OPE dictionary},''
  \href{http://dx.doi.org/10.1007/JHEP07(2022)079}{{\em JHEP} {\bfseries 07}
  (2022) 079}, \href{http://arxiv.org/abs/2110.04364}{{\ttfamily
  arXiv:2110.04364 [hep-th]}}.

\bibitem{Buric:2021ywo}
I.~Buric, S.~Lacroix, J.~A. Mann, L.~Quintavalle, and V.~Schomerus, ``{Gaudin
  models and multipoint conformal blocks: general theory},''
  \href{http://dx.doi.org/10.1007/JHEP10(2021)139}{{\em JHEP} {\bfseries 10}
  (2021) 139}, \href{http://arxiv.org/abs/2105.00021}{{\ttfamily
  arXiv:2105.00021 [hep-th]}}.

\bibitem{Buric:2020dyz}
I.~Buric, S.~Lacroix, J.~A. Mann, L.~Quintavalle, and V.~Schomerus, ``{From
  Gaudin Integrable Models to $d$-dimensional Multipoint Conformal Blocks},''
  \href{http://dx.doi.org/10.1103/PhysRevLett.126.021602}{{\em Phys. Rev.
  Lett.} {\bfseries 126} no.~2, (2021) 021602},
  \href{http://arxiv.org/abs/2009.11882}{{\ttfamily arXiv:2009.11882
  [hep-th]}}.

\bibitem{Antunes:2021kmm}
A.~Antunes, M.~S. Costa, V.~Goncalves, and J.~V. Boas, ``{Lightcone bootstrap
  at higher points},'' \href{http://dx.doi.org/10.1007/JHEP03(2022)139}{{\em
  JHEP} {\bfseries 03} (2022) 139},
  \href{http://arxiv.org/abs/2111.05453}{{\ttfamily arXiv:2111.05453
  [hep-th]}}.

\bibitem{Fortin:2022grf}
J.-F. Fortin, S.~Hoback, W.-J. Ma, S.~Parikh, and W.~Skiba, ``{Feynman rules
  for scalar conformal blocks},''
  \href{http://dx.doi.org/10.1007/JHEP10(2022)097}{{\em JHEP} {\bfseries 10}
  (2022) 097}, \href{http://arxiv.org/abs/2204.08909}{{\ttfamily
  arXiv:2204.08909 [hep-th]}}.

\bibitem{Rastelli:2017ymc}
L.~Rastelli and X.~Zhou, ``{Holographic Four-Point Functions in the (2, 0)
  Theory},'' \href{http://dx.doi.org/10.1007/JHEP06(2018)087}{{\em JHEP}
  {\bfseries 06} (2018) 087}, \href{http://arxiv.org/abs/1712.02788}{{\ttfamily
  arXiv:1712.02788 [hep-th]}}.

\bibitem{Behan:2021pzk}
C.~Behan, P.~Ferrero, and X.~Zhou, ``{More on holographic correlators: Twisted
  and dimensionally reduced structures},''
  \href{http://dx.doi.org/10.1007/JHEP04(2021)008}{{\em JHEP} {\bfseries 04}
  (2021) 008}, \href{http://arxiv.org/abs/2101.04114}{{\ttfamily
  arXiv:2101.04114 [hep-th]}}.

\bibitem{Belitsky:2014zha}
A.~V. Belitsky, S.~Hohenegger, G.~P. Korchemsky, and E.~Sokatchev, ``{N=4
  superconformal Ward identities for correlation functions},''
  \href{http://dx.doi.org/10.1016/j.nuclphysb.2016.01.008}{{\em Nucl. Phys. B}
  {\bfseries 904} (2016) 176--215},
  \href{http://arxiv.org/abs/1409.2502}{{\ttfamily arXiv:1409.2502 [hep-th]}}.

\end{thebibliography}\endgroup
